\documentclass[prd,twocolumn,showpacs,nofootinbib,superscriptaddress]{revtex4}

\usepackage[usenames]{color}
\usepackage{amsfonts}
\usepackage{amsmath}
\usepackage{amssymb}
\usepackage{bm}
\usepackage{dcolumn}
\usepackage{enumerate}
\usepackage{epsfig}
\usepackage{graphicx}
\usepackage{graphics}
\usepackage[latin1]{inputenc}
\usepackage{latexsym}
\usepackage{rotating}
\usepackage{hyperref}

\newcommand{\<}{\begin{equation}}
\newcommand{\?}{\end{equation}}

\newcommand{\R}{\mathbb{R}}
\newcommand{\Z}{\mathbb{Z}}

\newcommand{\C}{\mathbb{C}}

\newcommand{\Fc}{\mathfrak{c}}

\newcommand{\cA}{\mathcal{A}}
\newcommand{\cB}{\mathcal{B}}
\newcommand{\cC}{\mathcal{C}}
\newcommand{\cD}{\mathcal{D}}
\newcommand{\cE}{\mathcal{E}}
\newcommand{\cH}{\mathcal{H}}
\newcommand{\cR}{\mathcal{R}}

\DeclareMathOperator{\Real}{Re}

\newcommand{\eff}{\mathrm{eff}}
\newcommand{\hex}{\mathrm{hex}}
\newcommand{\lat}{\mathrm{lat}}
\newcommand{\cell}{\mathrm{cell}}
\newcommand{\even}{\mathrm{even}}
\newcommand{\odd}{\mathrm{odd}}
\newcommand{\QCD}{\mathrm{QCD}}
\newcommand{\es}{\mathrm{es}}
\newcommand{\iso}{\mathrm{iso}}

\newcommand{\Msolar}{M_{\odot}}

\begin{document}

\title{Shear modulus of the hadron--quark mixed phase}

\author{Nathan~K.~Johnson-McDaniel}

\affiliation{Institute for Gravitation and the Cosmos,
  Center for Particle and Gravitational Astrophysics,
  Department of Physics, The Pennsylvania State University, University
  Park, PA 16802, USA}
 \affiliation{Theoretisch-Physikalisches Institut,
  Friedrich-Schiller-Universit{\"a}t,
  Max-Wien-Platz 1,
  07743 Jena, Germany}

\author{Benjamin~J.~Owen}

\affiliation{Institute for Gravitation and the Cosmos,
  Center for Particle and Gravitational Astrophysics,
  Department of Physics, The Pennsylvania State University, University
  Park, PA 16802, USA}

\date{\today}

\begin{abstract}

Robust arguments predict that a hadron--quark mixed phase may exist in the
cores of some ``neutron'' stars.
Such a phase forms a crystalline lattice with a shear modulus higher than that
of the crust due to the high density and charge separation, even allowing for
the effects of charge screening.
This may lead to strong continuous gravitational-wave emission from rapidly
rotating neutron stars and gravitational-wave bursts associated with magnetar
flares and pulsar glitches.
We present the first detailed calculation of the shear modulus of the mixed
phase.
We describe the quark phase using the bag model plus first-order quantum
chromodynamics corrections and the hadronic phase using relativistic
mean-field models with parameters allowed by the most massive pulsar.
Most of the calculation involves treating the ``pasta phases'' of the lattice
via dimensional continuation, and we give a general method for computing
dimensionally continued lattice sums including the Debye model of charge
screening.
We compute all the shear components of the elastic modulus tensor and angle average
them to obtain the
effective (scalar) shear modulus for the case where the mixed phase is a
polycrystal.
We include the contributions from changing the cell size, which are necessary
for the stability of the lower-dimensional portions of the lattice.
Stability also requires a minimum surface tension, generally 
tens of MeV\,fm$^{-2}$, but dependent on the equation of state.
We find that the shear modulus can be a few times $10^{33}$~erg\,cm$^{-3}$,
two orders of magnitude higher than the first estimate, over a significant
fraction of the maximum mass stable star for certain parameter choices.

\end{abstract}


\maketitle

\section{Introduction}


Asymptotic freedom implies that matter at asymptotically high densities
consists of deconfined quarks, and these densities may overlap with the range
found in neutron stars.\footnote{We use the term ``neutron star'' to refer to any compact star made
of cold-catalyzed matter, regardless of actual composition.}
This was first noted by \citet{CoPe}, although before the discovery of
asymptotic freedom less concrete suggestions were made by \citet{Itoh} and
\citet{Bodmer}, and the first astrophysically concrete treatment (realistic
maximum mass, etc.)\ was given later by \citet{Witten1984}. More recent
developments include the realization that any quark matter in neutron stars
may be in a color superconducting state. The most well-known such state is the so-called
color-flavor locked (CFL) superconducting state present at asymptotically high densities~\cite{ARW}, but
there are other color superconducting states that may be present at lower densities---see~\cite{Alford2008} for a general review.
Quark matter could thus have a crystalline structure with a
very high shear modulus~\cite{MRS}, which became relevant to
gravitational-wave searches several years ago~\cite{Luca, LIGO_Crab}.

Even if none of these mechanisms applies and quark matter itself is fluid,
there is a generic and robust argument that neutron-star cores are filled with
a mixed phase of hadron and quark matter, which has a 
crystalline structure due to electrostatic and surface-tension effects.
Some recent gravitational-wave searches~\cite{LIGO_CasA, LIGO_magnetar} have
reached sensitivities for which the first rough estimate of the shear modulus
of this mixed phase~\cite{OwenPRL}---which is smaller than the estimates for
quark matter---is relevant.
Therefore it is worth a more careful calculation to get a better idea of which
gravitational-wave searches become relevant to this broader range of
theoretical models.
The justification for the existence of the mixed phase is as follows:

Glendenning~\cite{Glendenning91, Glendenning92} (see also
\cite{GlendenningPhR, Glendenning} for reviews) argued that the phase
transition from hadrons to quarks happens gradually, with a mixed phase
existing over a wide range of pressures.
A neutron star containing such a mixed phase is called a hybrid star.
The argument for this possibility is very robust (provided that the phase transition is
first order, as is generally expected), relying on the fact that
charge can be locally separated between the two phases while maintaining
global neutrality; this is energetically favored because it allows
the hadron phase to reduce its isospin asymmetry without producing leptons.
Observational constraints, even the discovery of a $1.97\pm0.04\,M_\odot$
pulsar~\cite{Demorestetal}, do not rule out the possibility that large regions
of the most massive neutron stars are composed of the mixed phase:
The input parameters of the models have more than enough uncertainty to
allow high-mass stars~\cite{Ozel2010, KBW, Weissenbornetal, CBBS, BoSe}.
(One can even still obtain pure quark cores in such massive stars, with appropriate
parameter choices~\cite{Weissenbornetal}, though we do not consider such
extreme cases here, for simplicity.)

Like the nuclear ``pasta'' phases in the crust (first discussed by Ravenhall, Pethick, and
Wilson~\cite{RPW}), the hadron--quark mixed phase is thought to form a
lattice.
This consists of charged blobs of the rare phase in a background of the common
phase and has
a varying dimensionality due to the competition between surface tension and the
Coulomb force, as first suggested by Heiselberg, Pethick, and Staubo~\cite{HPS}.
At the lowest pressure of the phase transition, small quark droplets form a
3-dimensional lattice in a background of hadrons.
At higher pressures the droplets grow and give way to a 2-dimensional lattice
of rods, then a 1-dimensional lattice of interleaved quark and hadron slabs.
At even higher pressures the progression is reversed, with the quark slabs
outgrowing the hadron slabs, then hadrons forming rods and droplets in a quark
background before vanishing altogether.

These mixed-phase lattices can have a larger shear modulus than that of the
crust, due mainly (in three dimensions) to the larger charge separation (from
several hundreds to more than a thousand elementary charges per blob, rather
than tens for nuclei).
The shear modulus was first estimated very roughly by one of
us~\cite{OwenPRL}, including a simple model of charge screening, the effects of which can
change the result by orders of magnitude.
A more detailed calculation was begun by~\citet{Nayyar}, and a rough estimate
neglecting charge screening but including the surface energy is given in
Sec.~7.7.2 of~Haensel, Potekhin, and Yakovlev~\cite{HPY}.
The latter also summarizes related work on other exotic phases:
High shear moduli for pion condensates have been predicted as early as
Ref.~\cite{PPS}, and similar estimates were made even earlier for solid
neutron cores which were proposed to explain glitches of the Vela
pulsar~\cite{PSR}.

Due to the high shear modulus of the mixed phase, hybrid stars could sustain
much larger deformations than normal neutron stars.
This has implications not only for pulsar glitches but also for
gravitational-wave emission, both continuous and in bursts associated with
magnetar flares~\cite{OwenPRL}.
Upper limits on the gravitational-wave energy emitted in magnetar flares have
entered the range of theoretical predictions (up to $10^{49}$\,erg for hybrid
stars)~\cite{Luca, LIGO_SGR, LIGO_SGRstacked, LIGO_magnetar}.
And the most interesting upper limits on continuous-wave emission, those that beat
the indirect limits (see \cite{Owen2009, Pitkin} for reviews), are for several stars
\cite{LIGO_Crab, LIGO_psrs2010, LIGO_CasA} within an order of magnitude of the
$10^{-5}$ maximum ellipticity first estimated for hybrid stars~\cite{OwenPRL}
and at the level of the $10^{-4}$ maximum that applies if more recent results
on the maximum crust breaking strain~\cite{HK} hold for the mixed
phase.
Therefore, more detailed calculations of the crystalline structure of the mixed
phase are interesting for the interpretation of gravitational-wave
observations even now, and will become more so when the upgraded ``advanced''
detectors come on-line.

Here we improve upon previous estimates of the mixed-phase shear
modulus~\cite{OwenPRL, Nayyar, HPY} with the first detailed calculation.
The implications for magnetar flares were discussed by \citet{CO}.
We will present an improved calculation of the consequences for continuous
waves elsewhere~\cite{paper2}.

We use the standard models used by \citet{Glendenning} for the hadronic and
quark equations of state (EOSs), as well as the standard Gibbs method for
calculating the bulk properties and lattice structure of the mixed phase.
We thus use a relativistic
mean field model for the hadronic matter. As Norsen and Reddy~\cite{NR} note, this assumption
can be of dubious validity for small regions of hadronic matter, since the fields do not take on their mean values in
that case.
However, this is not much of a concern for us, since, as Norsen and Reddy discuss, it is only a large effect for lower-dimensional
hadronic blobs, and these are only present in stable stars for one of the EOSs we consider, where
they have relatively large dimensions. (See Sec.~\ref{matter_models} for further discussion.) We do not have to worry about such errors in our description of the quark blobs, since we are using an improved bag model for them.
We also include the oft-neglected contribution from the surface tension to the
pressure balance in a few of our EOSs, following~\cite{NR,VYT_PLB,VYT_NPA,Endo}.

We obtain significantly larger shear moduli than Ref.~\cite{OwenPRL} primarily
due to our inclusion of the contributions that arise from the change in the cell
volume when one shears the lower-dimensional lattices.  (Some of these contributions were
considered by Pethick and Potekhin~\cite{PP} for the case of the nuclear pasta in the crust.
Additionally, the estimate of Haensel, Potekhin, and Yakovlev~\cite{HPY} is
based on these contributions.)
This is the most important improvement
in our analysis. These contributions are necessary for the lower-dimensional lattices to be
stable, and this stability leads to an EOS-parameter-dependent minimum surface
tension. Additionally, for large but reasonable surface tensions, these contributions significantly
increase the overall shear modulus of the lower-dimensional lattices.

Another fundamental improvement is that we treat the varying dimensionality of the
pasta phases using dimensional continuation.
This technique was first suggested by Ravenhall, Pethick, and Wilson~\cite{RPW} for the nuclear pasta in the
crust, and then applied to the hadron--quark pasta by \citet{GlendenningPhR}.
It is a relatively simple way to approximate the complicated structure seen in, e.g.,
molecular dynamics simulations of nuclear
pasta~\cite{Sonodaetal, Watanabeetal, Horowitzetal, HB}.
In order to perform the dimensionally continued lattice sums, we have
developed a generalization of the standard Ewald method~\cite{Ewald}
(see~\cite{JR} for a modern treatment).
The Ewald method was used by Fuchs~\cite{Fuchs} in his pioneering calculation
of elastic coefficients, and has very recently been used by Baiko~\cite{Baiko}
in a calculation of the effective shear modulus of the neutron star crust.
The underlying dimensionally continued Poisson summation formula has been
reported elsewhere by one of us~\cite{NKJ-M}.
Here we describe the practical implementation to the situation at hand,
including the details of the ``Ewald screening function'' and the dimensional
continuation of the specific family of lattices we consider.

We would
have liked to dimensionally continue the family of root lattices, $A_d^*$,
which solve the covering problem in dimensions $d\leq 5$ (see, e.g., Sec.~1.5 in Conway and Sloane~\cite{CS_lat}). However, as discussed
in Sec.~\ref{lat_dim_cont}, there is no
obvious way of doing so. Instead, we split the lattice up into hyperlattices
and introduce a freely specifiable interpolation
function for the hyperlattices' separation. We find that our results are quite
insensitive to the specific choice of
this function (particularly given the much larger uncertainties in the input
parameters). 

In yet another improvement, we perform rather than guess the angle averaging to
obtain a scalar shear modulus from the (anisotropic) elastic modulus tensor.
This makes the assumption (standard in the literature) that there are many
regions with random orientations, so that the mixed phase can be treated as a
polycrystal.
The magnetic fields present in a neutron star may cause this assumption to be violated, and relaxing
it would be an interesting topic for further investigation.

We also use the full Debye (linear) model for charge screening rather than
multiply the unscreened result by a simple correction factor.
Even this is a simplified treatment of screening effects, particularly because it is
linear.
Also, because we treat the blobs as point charges, it means that we do not include the
contribution of charge screening in our computation of the blobs' energy (which is used to
obtain the blob size and lattice spacing).
However a more detailed treatment (following,
e.g.,~\cite{NR,VYT_PLB,VYT_NPA,Endo}) would be much more computationally
intensive than the remainder of our approach, so we have left such
investigations to future work.
(Recent work by Endo~\cite{Endo} shows that the mixed phase can still occupy
much of the star even with a nonlinear treatment of charge screening, though
one cannot draw any definite conclusions about our results from Endo's because
he uses different EOS parameters.)
Some indication of the magnitude of the error we make in using the Debye model
can be seen in the jumps in the shear modulus in the figures in
Sec.~\ref{results1}.

We use electrostatic units and set $\hbar = c = 1$, so we will generally
express masses in MeV and lengths in fm. All the computations were performed using
{\sc{Mathematica}}~7's default methods to solve algebraic equations, numerically evaluate integrals,
etc.

The paper is structured as follows:
In Sec.~\ref{matter_models} we review the models we use for the hadronic and
quark EOSs, as well as how we compute the bulk properties, lattice
structure, and charge screening of the mixed phase.
In Sec.~\ref{elasticity}, we give an overview of the elastic response of the
lattice for the various
integer dimensions, and show how we compute an average shear modulus
in a dimensionally continued manner. We describe how to compute the relevant
elastic constants in Sec.~\ref{elastic_calc}, and how to dimensionally continue the resulting
lattice sums in Sec.~\ref{dim_cont}. We give our results for the shear modulus in Sec.~\ref{results1}, along with some discussion,
and conclude in Sec.~\ref{concl1}. We
discuss various checks on our computations of the lattice sums in the
Appendix.

\section{Mixed phase}
\label{matter_models}

We model the hadronic and quark phases following
Glendenning~\cite{Glendenning}, using a relativistic mean field theory model for
the hadronic matter (from Glendenning's Chap.~4), and an improved bag model description of the quark matter (from Glendenning's Chap.~8). (For simplicity, we do not consider the possibility of
color superconductivity.)
We have chosen to compute the EOSs ``from scratch'' instead of using any sort of
tabulated EOS (except for the low-density EOS, which has a negligible effect on our results).
This allows us to include the effects of surface tension on the EOS
(as discussed in Sec.~\ref{lat_struc}), and
to investigate the effects of different EOS parameters on the shear modulus.
It also lets us compute the Debye
screening length, for which we need to know how
the chemical potentials of the particle species vary with density.
The models we use for the hadronic and quark phases are relatively simple, but should contain at least a rough
description of the relevant physics for the situation under consideration. And given the significant uncertainties associated with any description of
cold, dense matter, even relatively simple models should allow an adequate sampling of the
relevant parameter space.

As mentioned previously---and emphasized by Norsen and Reddy~\cite{NR}---the mean field theory description is not accurate for small regions of hadronic matter.
However, this is of little concern for us, since this should only be a large effect for small, low-dimensional hadronic blobs, as discussed by Norsen and Reddy.
The only EOS for which one obtains such blobs in stable stars,
LKR$1$, has low-dimensional hadronic blob radii of $\sim 10$~fm, as illustrated in Sec.~\ref{lat_struc}. Taking the hadron-kaon
results from Norsen and Reddy to apply to our case, we see that the largest finite-size corrections
will thus be $\sim10\%$.

These EOSs depend on a variety of parameters, discussed in more detail in the
following subsections. We consider a small
collection of
representative sets of parameters given in
Table~\ref{EOS_table}.
All of these parameter sets are chosen to yield a
maximum Oppenheimer-Volkov (OV) mass compatible with the recent
\citet{Demorestetal} measurement of a $1.97\pm 0.04 \Msolar$ neutron star.

Since the hadronic EOS we consider is only valid at densities
well above neutron drip, we add on a standard low-density EOS for baryon
number densities $n_B < 0.08\text{ fm}^{-3}$.
This is the combination of the Baym, Pethick, and Sutherland (BPS)~\cite{BPS} EOS
for $n_B < 0.001\text{ fm}^{-3}$ and
the \citet{NV} EOS for $0.001\text{ fm}^{-3} < n_B < 0.08\text{ fm}^{-3}$
used by \citet{LP2001}.
We use the table provided by
Kurkela \emph{et al.}~\cite{Kurkelaetal} at~\cite{Kurkelaetal_URL}.
The precise choice of the low-density part of the EOS has a negligible effect
on the maximum mass of a stable star.

Since we are interested in how large astrophysical effects could be, we generally choose our
EOS parameters to be middle-ground estimates, but also consider some more
extreme cases, to yield massive stars with a large region of mixed phase. (However, we do not consider
stars with quark cores, for simplicity.)
The Hy1 EOS is taken from \citet{Nayyar}, correcting an error in his code
which led to a lower maximum mass.
The Hy$1'$ EOS changes the
quark bag constant to obtain a larger region of mixed phase with
more pasta phases in stable stars. The Hy$1\mu$ and Hy$1\sigma$ EOSs each
change the treatment of one portion of the calculation to give some indication
of how much these affect our results: Hy$1\mu$ uses a
chemical-potential-dependent quantum chromodynamics (QCD) renormalization scale,
while Hy$1\sigma$ includes the effects of surface tension on the pressure
balance of the two phases; Hy$1\mu\sigma$ includes both.

Following
Weissenborn \emph{et al.}~\cite{Weissenbornetal}, we also consider a case called LKR1 with
the NL3 hadronic EOS parameters of Lalazissis, K{\"{o}}nig, and Ring  (LKR)~\cite{LKR}. These parameters give a very
stiff hadronic EOS (a purely hadronic star would have a maximum mass of
$2.78\Msolar$) and thus allow for neutron stars with a large region of
mixed phase that
still are compatible with a $1.97\pm0.04\,M_\odot$ neutron star.
For the corresponding quark EOS we picked parameters that lead to stable stars that include all
the pasta phases.
We also considered a case with somewhat generic parameters, generally picking
them to be around the midpoint of the accepted range---this is the ``generic''
parameter set. These were not fine-tuned at all.
Additionally, we have considered a variant that uses these more modern
hadronic EOS parameters along with a bag constant and QCD coupling constant
that yield a larger region of mixed phase---this is the generic${}'$ parameter
set.

\begin{figure}[htb]
\begin{center}
\epsfig{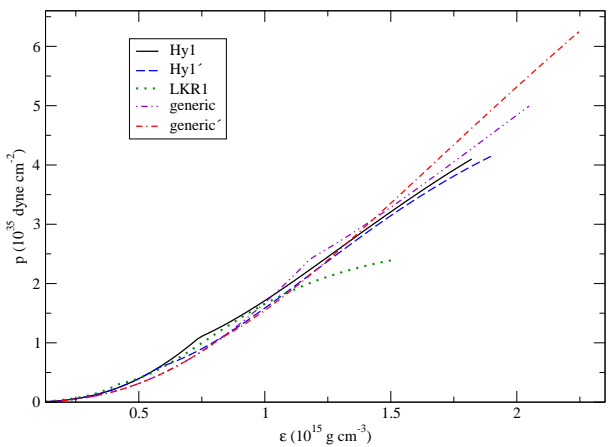}
\end{center}
\caption[Pressure vs.\ energy density for the EOSs from
Table~\ref{EOS_table}]{\label{EOSs} Pressure vs.\ energy density for the EOSs from
Table~\ref{EOS_table}, plotted up to the maximum energy density present in a 
stable OV star.}
\end{figure}

\begin{figure}[htb]
\begin{center}
\epsfig{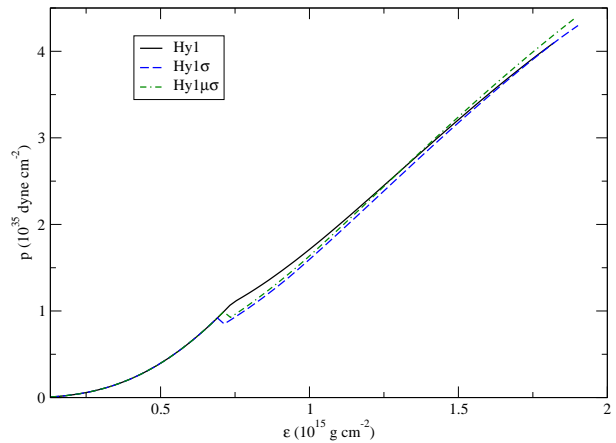}
\end{center}
\caption[Pressure vs.\ energy density for the Hy$1$, Hy$1\sigma$, and Hy$1\mu\sigma$ EOSs]{\label{Hy1_flavours} Pressure vs.\ energy density for the Hy$1$ EOS
and two flavors that include the surface tension contribution to the pressure
balance, illustrating the violations of le Chatelier's principle. (Again, each
of these is plotted up to the maximum energy density present in a 
stable OV star.)}
\end{figure}

We plot these EOSs up to the maximum density present in a stable OV star in
Fig.~\ref{EOSs}. We do not show the various flavors of the Hy$1$ EOS here,
since the resulting traces are all very similar, though we do show the small violations of le Chatelier's
principle for the Hy$1\sigma$ and Hy$1\mu\sigma$ EOSs, compared with the
Hy$1$ EOS in Fig.~\ref{Hy1_flavours}. (We do not show the trace for the
Hy$1\mu$ EOS, as it is almost identical to that for the Hy$1$ EOS.)

\begin{table*}
\begin{tabular}{cccccccccccccccc}
\hline\hline
 & $n_0$ & $(E/A)_\infty$ & $K$ & $m^*/m$ & $a_\mathrm{sym}$ & $B^{1/4}$ & $\alpha_s$ & $\bar{\Lambda}_\QCD$ & $m_s$ & $\sigma$ & $M_\mathrm{max}$ & $M^\mathrm{hybrid}_\mathrm{min}$ & $R^\mathrm{hybrid}_\mathrm{max}/R$ & $\cC_\mathrm{max}$ & densest\\
 & $\text{fm}^{-3}$ & MeV & MeV & & MeV & MeV & & MeV & MeV & $\text{MeV fm}^{-2}$ & $\Msolar$ & $\Msolar$ & $\%$ & & hybrid phase\\
\hline
Hy1 & $0.153$ & $-16.3$ & $300\phantom{.00}$ & $0.7\phantom{00}$ & $32.5$ & $180\phantom{.0}$ & $0.6$ & $300$ & $150$ & $10$, $20$, $50$, $80$ & $2.057$ & $1.747$ & $57.7$ & $0.484$ & Q, $d = 1.03$\\
Hy$1\mu$ & $0.153$ & $-16.3$ & $300\phantom{.00}$ & $0.7\phantom{00}$ & $32.5$ & $180\phantom{.0}$ & $0.6$ & $\bar{\mu}_q$ & $150$ & $80$ & $2.089$ & $1.794$ & $56.6$ & $0.490$ & Q, $d = 1.23$\\
Hy$1\sigma$ & $0.153$ & $-16.3$ & $300\phantom{.00}$ & $0.7\phantom{00}$ & $32.5$ & $180\phantom{.0}$ & $0.6$ & $300$ & $150$ & $80$, p$\negthickspace\negthickspace\negthickspace\negthinspace\negthinspace$ & $1.997$ & $1.583$ & $63.6$ & $0.481$ & Q, $d = 1.00$\\
Hy$1\mu\sigma$ & $0.153$ & $-16.3$ & $300\phantom{.00}$ & $0.7\phantom{00}$ & $32.5$ & $180\phantom{.0}$ & $0.6$ & $\bar{\mu}_q$ & $150$ & $80$, p$\negthickspace\negthickspace\negthickspace\negthinspace\negthinspace$ & $2.025$ & $1.660$ & $61.7$ & $0.486$ & Q, $d = 1.05$\\
Hy$1'$ & $0.153$ & $-16.3$ & $300\phantom{.00}$ & $0.7\phantom{00}$ & $32.5$ & $170\phantom{.0}$ & $0.6$ & $300$ & $150$ & $80$ & $1.974$ & $1.377$ & $69.0$ & $0.476$ & H, $d = 1.30$ \\
LKR1 & $0.148$ & $-16.3$ & $271.76$ & $0.6\phantom{00}$ & $37.4$ & $167.5$ & $0.6$ & $300$
& $100$ & $80$ & $1.955$ & $1.096$ & $72.5$ & $0.433$ & H, $d = 3.00$\\
generic & $0.16\phantom{0}$ & $-16\phantom{.0}$ & $250\phantom{.00}$ & $0.745$ & $30\phantom{.0}$ & $200\phantom{.0}$ & $0.5$ & $\bar{\mu}_q$
& $100$ & $80$ & $1.986$ & $1.878$ & $44.0$ & $0.500$ & Q, $d = 2.10$\\
generic${}'$ & $0.16\phantom{0}$ & $-16\phantom{.0}$ & $250\phantom{.00}$ & $0.745$ & $30\phantom{.0}$ & $170\phantom{.0}$ & $0.7$ & $\bar{\mu}_q$
& $100$ & $80$ & $1.974$ & $1.534$ & $65.9$ & $0.515$ & Q, $d = 1.36$\\
\hline\hline
\end{tabular}
\caption[EOS parameters]{\label{EOS_table} EOS parameters and properties of their associated OV stars.
In the $\bar{\Lambda}_\QCD$
column, $\bar{\mu}_q$ denotes the cases where we take the QCD renormalization
scale to be given by the average quark chemical potential at each density, as
discussed in
Sec.~\ref{quarkEOS}. In the $\sigma$ column, we have denoted the cases in
which we include the surface tension contribution to the pressure balance by a
``p.'' $M^\mathrm{hybrid}_\mathrm{min}$ gives the masses of stars that first contain hybrid matter; $R^\mathrm{hybrid}_\mathrm{max}/R$ denotes the maximum radius fraction occupied by hybrid matter
(i.e., the radius fraction for the maximum mass star); and $\cC_\mathrm{max}$
denotes the maximum compactness ($2GM/Rc^2$) of a star. We also give the
composition of the rare phase (``Q'' for quark and ``H'' for hadronic) and the
dimension of the lattice at the center of the maximum mass star.
See the text for the definitions of other parameters.
}
\end{table*}

\subsection{Hadronic EOS}

We construct the hadronic EOS following the recipe described in
Chap.~4 of Glendenning's book~\cite{Glendenning} through Sec.~4.9.
We thus use
a relativistic mean field description, with a Lagrangian that contains the
standard scalar and vector ($\sigma$ and $\boldsymbol{\omega}$) fields, as well as scalar
self-interactions, and the isospin asymmetry force, mediated by the vector
meson $\boldsymbol{\rho}$. We include neutrons, protons, electrons, and muons. See~\cite{Nayyar} and~\cite{LNO} for further details about the general framework and calculational procedure we
used. However, we have used updated parameters in the models, as discussed
above. Note that these references also consider hyperons, which we do not include in the EOSs considered here. In particular, hyperons and the hadron--quark mixed phase have been found to be mutually exclusive in
some studies (see, e.g.,~\cite{CBBS, MCST}).

The model has five input
parameters: Two are known reasonably well, viz., the number density and
binding energy per nucleon of nuclear matter at saturation,
$n_0 = 0.16 \pm 0.01 \text{ fm}^{-3}$ and $(E/A)_\infty =-16\pm1\text{ MeV}$.
The remaining three 
are not nearly so well known: The nuclear
incompressibility $K$ is thought to lie between $200$ and $300$~MeV, with
many authors placing it in the range $240\pm10$~MeV (see~\cite{Piekarewicz}),
while
the (scaled) Dirac effective mass of the nucleons, $m^*/m$, is thought to be
between $\sim 0.53$
and $0.96$. (Here $m = 938.93 \text{ MeV} = 4.7582 \text{ fm}^{-1}$ is the average of the neutron and
proton masses.) The symmetry energy $a_\mathrm{sym}$ is thought to lie between $28$ and $34$~MeV (see Li~\emph{et al.}~\cite{Lietal}).

All parameter ranges are taken from Eq.~(88) in
Steiner~\emph{et al.}~\cite{Steiner_etal} unless otherwise noted. The range for
the Dirac
effective mass is computed from that for the Landau effective mass given by
Steiner~\emph{et al.}, following Glendenning's Eq.~(4.117)~\cite{Glendenning},
with a Fermi momentum at saturation of $1.33 \pm 0.03 \text{ fm}^{-1}$, from
the Steiner~\emph{et al.}\
result and Glendenning's Eq.~(4.110).

Note that almost all of the NL$3$ parameters from LKR fall outside of the ranges
we have given. We still consider them for two reasons. First, the LKR NL$3$ EOS is often
treated as the paradigmatic stiff hadronic EOS in the literature; in
particular, we were inspired to consider these parameters by the LKR EOS's
recent use in
Weissenborn~\emph{et al.}~\cite{Weissenbornetal}. Second, the LKR NL$3$ parameters
are still regarded as providing an excellent fit to the nuclear binding energy and
charge radius, explaining why this EOS remains in use.

\subsection{Quark EOS}
\label{quarkEOS}

The quark matter is described by the MIT bag model (first given in Chodos
\emph{et al.}~\cite{Chodosetal}, and discussed briefly in Glendenning~\cite{Glendenning}),
to capture the basic physics of confinement.
This is supplemented with first-order QCD
corrections to the thermodynamic potential for free quarks from Farhi and
Jaffe~\cite{FJ}, as described in Chap.~8 of Glendenning~\cite{Glendenning}.\footnote{Note that Glendenning corrects a sign error in Farhi and Jaffe's result for
the thermodynamic potential in his Eq.~(8.14).} We take
all the quarks to be massive (as opposed to the usual treatment in which only
the strange quark is taken to be massive---see, e.g.,~\cite{Alfordetal, KRV,
Weissenbornetal}---but note that Christiansen and Glendenning~\cite{ChGl97} also take all three quarks to be
massive). 
See Chap.~6 in~\cite{Nayyar} for further details of the calculation.

Here the parameters are all quite uncertain. The only firm constraint is that
nonstrange quark matter is not absolutely stable, since this contradicts the
observed existence of nuclei composed of nucleons. In addition, we assume that
strange quark matter is not absolutely stable, since we consider hybrid stars,
not strange stars. In the pure bag model (with no QCD corrections---i.e., with
zero QCD coupling constant), this
implies that the fourth root of the bag constant, $B^{1/4}$, is greater than $145$~MeV. When one
includes QCD corrections, the minimum bag constant decreases---see, e.g., the
discussion in Alford~\emph{et al.}~\cite{Alfordetal}.
There is no known upper bound on $B$, though for sufficiently large values of
$B$ (with all other EOS parameters fixed), stable stars do not contain
deconfined quark matter.

We also need to
know the QCD coupling constant and quark masses at the energy scale $\bar{\Lambda}_\QCD$ at which we renormalize the
perturbative contributions to the thermodynamic potential. This energy scale is typically
taken to
be the scale of the quark chemical potentials in the problem. Glendenning uses
$300$~MeV, though we find that the chemical potentials in the situations we
consider are somewhat higher (up to $\sim 480$~MeV in the densest regions of
stable stars).
Therefore we also consider a density-dependent renormalization scale,
given by $\bar{\Lambda}_\QCD = \bar{\mu}_q := (\mu_p+\mu_n)/6$, where $\mu_p$ and $\mu_n$ are the
proton and neutron chemical potentials, respectively, so $\bar{\mu}_q$ is an
average quark chemical potential. (This was inspired by the chemical
potential-dependent renormalization scale used in~\cite{KRV}, though the
specifics of their treatment differ.)

At this scale, the QCD coupling constant $\alpha_s$ is large and not well known
(see Fig.~1 in~\cite{KRV} for the results from running the coupling constant
using relatively low-order beta function expressions). We have
considered values between $0.5$ and $0.7$, which are quite middle-of-the-road
(particularly compared with the na{\"\i}ve beta function result of much
greater than $1$, though there is evidence that the QCD coupling constant ``freezes'' to a value of
less than $1$ at low
energies---see, e.g.,~\cite{MaSt, Ganbold}).
Weissenborn \emph{et al.}~\cite{Weissenbornetal} consider values from $0$ to
$0.94$,\footnote{This is converted from Weissenborn \emph{et al.}'s range of $0.4$ to $1$
for Alford \emph{et al.}'s $a_4$~\cite{Alfordetal} using the massless quark expression of
$\alpha_s = (1 - a_4)\pi/2$ [cf.\ Eq.~(3) in Alford \emph{et al.}~\cite{Alfordetal} and Eq.~(6.4) in Nayyar~\cite{Nayyar}~].}  while Nayyar~\cite{Nayyar} uses values of $0.45$ and $0.6$.
The quark masses are similarly
uncertain. There are significant uncertainties (up to $\sim 50\%$) even at the $2$~GeV scale at
which the Particle Data
Group~\cite{PDG} quotes the masses, and additional uncertainties in running them to lower energies.
For simplicity, we have taken the up, down, and strange quarks to have masses of $2.5$, $5$,
and $100$~MeV,
respectively, around the median of ranges for the $\overline{\mathrm{MS}}$
current masses at $2$~GeV given in the 2010 Review of Particle Physics~\cite{PDG}. We also
consider a few cases with a strange quark mass of $150$~MeV, for continuity with Nayyar~\cite{Nayyar}.

\subsection{Hybrid phase and its lattice structure}
\label{lat_struc}

We determine the bulk properties of the mixed phase using the Gibbs equilibrium
conditions, following Glendenning (see, e.g., Chap.~9 in~\cite{Glendenning}):
The appearance of the mixed phase at a certain baryon number density is
signaled by pure quark matter having a larger pressure than hadronic matter.
We then determine the volume fractions of the two coexisting phases by demanding the
equality of the phases' pressures and chemical potentials, in addition to
global electrical neutrality. We compute the mixed-phase bulk pressure, energy,
and baryon densities by weighting the contributions from each phase using their
volume fractions.

We also need to obtain the crystalline structure of the mixed phase.
This requires knowing the surface
tension $\sigma$ of the hadron--quark interface. While $\sigma$ has been estimated very roughly in the
literature (e.g.,~\cite{FJ} and the references given in Endo~\cite{Endo}), its value is still quite uncertain, so we simply take it to
range from $10$ to $80 \text{ MeV fm}^{-2}$. (The lower bound is the default
value used by Nayyar~\cite{Nayyar}, and the upper bound is the default value
used by Glendenning~\cite{Glendenning}.)
We make sure that it does not make an appreciable contribution
to the overall energy density, since, as Glendenning discusses, the system's energy should not
increase upon opening up a new degree of freedom. In the cases we have considered, the
blobs' energy density is $\lesssim 2\%$ of the total. 
Similarly, the lattice's electrostatic pressure is also $\lesssim3\%$ of the
bulk pressure.

Although the Coulomb and surface energy of the blobs is only a small contribution to the overall energy density, it can still be significant compared to the energy difference between the pure hadronic phase and the mixed phase at a given baryon density,
particularly at lower densities. Thus, as discussed by Heiselberg, Pethick, and Staubo~\cite{HPS} and Alford~\emph{et al.}~\cite{Alfordetal_PRD}, for sufficiently high surface tensions, the mixed phase may be disfavored 
compared to the sharp transition predicted by the Maxwell construction. Christiansen and
Glendenning, however, argue that the Maxwell construction should always correspond to an
excited state for a multi-component system such as we are considering, and if any models predict otherwise for certain parameter values, they are not accurate for those parameters~\cite{ChGl97,ChGl00}.

Regardless, these energy arguments are all local (i.e., at a fixed baryon density). If one considers
the energy of the star as a whole (at a fixed total baryon number), then the mixed phase can still be favored, particularly if one
accounts for the increase of the star's own gravitational binding energy due to the attendant
softening of the EOS at high densities. We shall
present our analysis of the extent to which the mixed phase is present in stable
stars for various surface tensions when we consider the associated maximum quadrupoles in~\cite{paper2}. 
Suffice it to say that given all the uncertainties present in these calculations, we still feel comfortable quoting results for surface tensions of $80 \text{ MeV fm}^{-2}$, at least as an indication of the upper bounds possible for the shear modulus. (We also discuss the dependence of the shear modulus on surface tension in Sec.~\ref{results1}.) Note that if one includes the lattice contributions to the EOS (energy
density and pressure, plus the energy density of the blobs themselves), one obtains surface tension-dependent corrections to the values for stellar properties given in Table~\ref{EOS_table}. However, these are at most a few percent for the surface tensions we consider here.

Returning to the lattice structure, we note that the competition between the aforementioned surface tension and the Coulomb energy of
the charges, the charge breaks up into blobs, which will then form a
lattice. In order
to determine the properties of this lattice, we employ the Wigner-Seitz
approximation, which divides the lattice up into noninteracting, electrically
neutral cells (approximating the lattice's Voronoi cell by a sphere of
equivalent volume).
Each cell contains a charged blob at its center
surrounded by an equal amount of compensating charge.
At lower densities, the
blob is formed out of quark matter, and the hadronic matter provides the
compensating charge; at higher densities---not reached in stable
stars for many EOSs---the roles are reversed. The ratios of the 
volumes of the blob and the cell are fixed, since the baryon and quark volume
fractions are fixed. If we denote the volume fraction of the rare
phase by $x$, then we have $x = (r/R)^d$, where
$r$ and $R$ are the radii of the blobs and cells, respectively (the
half-thickness of the slabs in the one-dimensional case), and $d$ is
the dimensionality of the cell and lattice. If we denote the volume fraction of
the quark phase by $\chi$, we have
\<\label{x_eq}
x = \begin{cases}
\chi, & \chi \leq 1/2,\\
1-\chi, & \chi> 1/2.
\end{cases}
\?
We then determine $r$ and $d$ by
minimizing the cell's energy per unit volume.

Following the suggestion initially made by Ravenhall, Pethick, and
Wilson~\cite{RPW} in the nuclear pasta case, we take the lattice's
dimension to be a continuous variable.
In the integer
dimension cases, we have $A_d^*$: a body-centered cubic (bcc) lattice of spherical drops in $3$ dimensions, a hexagonal lattice
of cylindrical rods in $2$ dimensions, and equally spaced rectangular
slabs in $1$ dimension. (These lattices are seen in the molecular dynamics
simulations of nuclear pasta~\cite{Sonodaetal,Watanabeetal,Horowitzetal, HB}.
In particular, Watanabe~\emph{et al.}~\cite{Watanabeetal}
obtain a hexagonal lattice of rods by adiabatically compressing a bcc lattice
of drops in their molecular dynamics simulations.)

\begin{figure}[htb]
\begin{center}
\epsfig{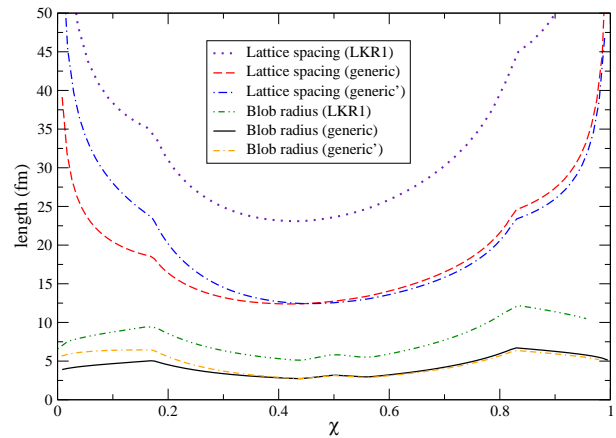}
\end{center}
\caption{\label{lattice_quantities} The blob radius and lattice spacing versus $\chi$ for three
representative EOSs.}
\end{figure}

The expression for the energy per unit volume of a
cell is then given by [Eqs.~(9.19), (9.23), and (9.24) in
Glendenning~\cite{Glendenning}~]
\<\label{cell_energy}
\frac{E_\cell}{\Omega} = C(x)r^2 + \frac{S(x)}{r},
\?
where $\Omega$ is the cell volume, and $C(x)$ and $S(x)$ correspond to the
Coulomb and surface contributions, respectively, with
\<
C(x) := 2\pi[(q_H - q_Q)(\chi)]^2xf_d(x),\quad S(x) := x\sigma d.
\?
Here $q_H$ and $q_Q$ are the hadron and quark
charge densities and
\<
f_d(x) := \frac{1}{d+2}\left[\frac{2 - dx^{1-2/d}}{d-2} + x\right].
\?
The singularity at $d \to 2$ is removable, since the limit exists [and
has the expected value; see Eq.~(9.30) in Glendenning~\cite{Glendenning}~].
One can now immediately read off the minimizing $r$ [Eq.~(9.27) in
Glendenning~\cite{Glendenning}~], viz.,
\<
r = \left[\frac{S(x)}{2C(x)}\right]^{1/3}.
\?
One then obtains $d\in[1,3]$ by minimizing (over that range) the resulting
expression for the cell energy (using the minimizing $r$), viz.,
$E_\cell/\Omega = (3/2)[2C(x)]^{1/3}[S(x)]^{2/3}$ [see Eq.~(9.28) in Glendenning~\cite{Glendenning} for an explicit expression in terms of $x$, etc.].
We show the dependence of blob radius and lattice spacing [given in Eq.~\eqref{scaling}] on
$\chi$ for a few representative EOSs in Fig.~\ref{lattice_quantities}. (The blob radii and lattice spacings of the three EOSs shown span the range covered by the entire set of EOSs in Table~\ref{EOS_table}.)

As discussed in, e.g.,~\cite{NR, VYT_PLB, VYT_NPA, Endo},
the surface tension contributes to the pressure equality. Explicitly, we have
a pressure difference between the dominant and rare phases of [cf.\ Eq.~(A5)
in~\cite{ChGl97}~]
\<
p_\mathrm{dominant} - p_\mathrm{rare} = \frac{(d-1)\sigma}{r},
\?
where $r$ is the radius of the blobs. We have included this in our treatment of
certain EOSs, though in a somewhat simplified version, so that it is numerically
tractable. One obtains these EOSs by solving the appropriate equations at progressively
larger and larger baryon densities (as discussed by Glendenning~\cite{Glendenning}). Our
simplification
consists of taking the lattice's dimension at a given baryon density to be fixed at the value
obtained at the previous baryon density (instead of solving for the dimension including the
dimension-dependent surface tension contribution to the pressure equality).
Since this correction to the pressure
equality is largest at the lowest-density portions of the mixed phase (i.e., the quark drop portions,
with $d=3$), our procedure for $d$ seems likely to account for the primary effects from the surface tension's
contribution to the pressure equality. (Note that we take the bulk pressure to be that of the dominant phase.)

With this treatment of the surface tension contribution to the pressure equality, we find a slight
violation of le Chatelier's principle (i.e., monotonic increase of pressure with
energy density) close to the hybrid transition.
This is not present if one does not include the surface
tension contribution to the pressure equality (see Fig.~\ref{Hy1_flavours}).
Note that our method
of solving the OV equations---using the enthalpy form given by Lindblom~\cite{Lindblom}---does not allow for such violations of le Chatelier's principle, since it requires one to express the energy density as a function of the pressure. We nonetheless quote results using the ``smoothed-out''
version of the EOS produced by {\sc{Mathematica}}'s interpolation, feeling that they still give a reasonably accurate depiction of the star's bulk properties, since the violations of le Chatelier's principle we
are considering are small.

\subsection{Charge screening}
\label{screening}

The previous discussion has assumed that the charge is uniformly distributed within each blob and in the
neutralizing background. This is not the case, in practice: The minimum-energy configuration will have a nonuniform charge distribution, as is
discussed in, e.g.,~\cite{NR, VYT_PLB, VYT_NPA}. This nonuniform charge
distribution---often treated in the perturbative regime as charge
screening---will affect the cell energy (and thus the lattice properties, for
a given energy density), as well as the lattice's electrostatic
energy, and both of these affect the shear modulus. Here we only consider the
effects on the lattice's electrostatic energy due to linear charge screening
of point charges, using a screened potential. As is commonly done in 
treatments of the mixed phase (e.g., those by Glendenning), we do not
concern ourselves at all with the rearrangement of charge inside the blobs,
though this is likely a significant effect (as is discussed
in~\cite{NR, VYT_PLB, VYT_NPA}).

In this linearized version, we treat screening as a small perturbation on the
overall energy, leading to the expression for the Debye length given
in Eq.~(1) of~\cite{HPS}, viz.,
\<
\lambda = \left[4\pi\sum_\alpha Q_\alpha^2\left(\frac{\partial n_\alpha}{\partial \mu_\alpha}\right)\right]^{-1/2}.
\?
Here $Q_\alpha$, $n_\alpha$, and $\mu_\alpha$ are the charge, number density, and chemical potential
of particle species $\alpha$, and the partial derivative is evaluated holding the chemical
potentials of all species besides the $\alpha$th fixed.

To include the effects of charge screening on the electrostatic
energy, we dimensionally continue the standard screened potential equation
(Yukawa in three dimensions), viz.,
\<\label{phiPDE}
(\triangle_d - \lambda^{-2})\phi = -4\pi Q\delta^{(d)},
\?
where $\triangle_d$ is the $d$-dimensional Laplacian,
$Q$ is the charge of the blob,
and $\delta^{(d)}$ is the $d$-dimensional Dirac delta distribution.

The resulting potential is
\<\label{phi}
\phi(r) = \frac{2Q}{(2\pi \lambda r)^{d/2-1}}K_{d/2 - 1}(r/\lambda),
\?
where $K_\nu$ is the modified Bessel function of the second kind of order
$\nu$.
This expression can be obtained most easily by noting that it is the same as
the Euclidean scalar propagator (up to overall factors), which is given in a
dimensionally continued form in Eq.~(5) of~\cite{PS}.\footnote{When comparing
the two expressions, note that
$K_{-\nu} = K_\nu$.}

This potential clearly represents a simplified treatment of charge screening even in the perturbative
limit, since we have used a point charge, instead of the extended charge distribution of a realistic
blob with different screening lengths inside and outside. (We similarly use
point charges when calculating the potential energy of other blobs due to this
screened potential.) However, it is a practical way for us to account for some
of the effects of charge screening. Since we are
using this point charge approximation, we compute the screening length using
the leptons and the background phase.
This is because the rearrangement of
charge within the blob has no effect on the blob's electrostatic potential except through the
effects of screening on the blob's size, which we are ignoring here. This
switch at $\chi = 1/2$ causes a significant jump in the screening length there. This
translates into a jump in the elastic constants, and thus the effective shear
modulus, as seen in the figures in Sec.~\ref{results1}. The magnitude of this
jump gives some indication of the overall error we incur through our simplified
treatment of charge screening in the potential. (However, this likely does not
give any indication of the errors incurred by neglecting the contribution of
charge screening to the cell energy, which we shall see affects the effective
shear modulus both directly and indirectly.)

\section{Elasticity}
\label{elasticity}

The elastic response of a crystalline lattice is described by its
elastic modulus tensor, $S_{klps}$, defined by
\<\label{Sklps}
\cE = \cE_0 + S_{kl}u_{kl} + \frac{1}{2}S_{klps}u_{kl}u_{ps},
\?
where $\cE$ is the deformed lattice's energy density, $\cE_0$ is its
undeformed energy density, $u_{kl} := \partial_k\delta x_l$ is the displacement gradient (where
$\delta x_l$ denotes the displacement field), and $S_{kl}$ is the
first-order piece of the expansion. We have a nonzero first-order piece here,
since the undeformed lattice has a nonzero pressure---see Eq.~(7) in
Baiko~\cite{Baiko}. (Wallace~\cite{Wallace} gives further discussion.)
Explicitly, we have $S_{kl} = -(P_\iso + P_\es)\delta_{kl}$ for a three-dimensional lattice, where $P_\iso$ is
the isotropic portion of the lattice's pressure (which we shall see does not contribute to the shear modulus, as expected), and $P_\es$ denotes the lattice's electrostatic pressure. (Of course, the elecrostatic pressure is also isotropic in three dimensions, but the portion that we are referring to as isotropic
contains the contributions from changing the cell volume and charge that will remain isotropic as the lattice dimension decreases; these contributions to the shear elastic constants cancel explicitly in the
three-dimensional case.) For a two-dimensional lattice, with
the $z$-direction perpendicular to the lattice, we have $S_{kl} = -(P_\iso + P_\es)\delta_{kl} + P_\es\hat{z}_{kl}$, and similarly, for a one-dimensional lattice with
the lattice spacing in the $x$-direction, we have $S_{kl} = -P_\iso\delta_{kl} - P_\es\hat{x}_{kl}$. (Here $\hat{z}_k$ denotes a unit vector in the $z$-direction, $\hat{z}_{kl} := \hat{z}_k\hat{z}_l$, and
similarly for $\hat{x}_{kl}$.)
Now, we are only interested in the shear stress the lattice generates in response to a
shear deformation, and this is given by a different elastic modulus tensor, viz.,
\<
\begin{split}
B_{klps} &= S_{klps} + \frac{1}{2}(T_{kp}\delta_{ls} + T_{ks}\delta_{lp} + T_{lp}\delta_{ks} - T_{ls}\delta_{kp}\\
&\quad - 2T_{kl}\delta_{ps}),
\end{split}
\?
where $T_{kl} = S_{kl}$ is the stress tensor
[Eqs.~(2.24) and~(2.36) in Wallace~\cite{Wallace}~].
Specifically, the stress generated by a deformation $u_{kl}$ is given by
$B_{klps}(u_{ps} + u_{sp})/2$. [See Eqs.~(29) and (30) in Baiko~\cite{Baiko}.] We can write the components
of $B_{klps}$ more simply by using Voigt notation to map them to the two-index object
$c_{\alpha\beta}$
using the index mapping given by
$\{xx,yy,zz,xy,xz,yz\}\leftrightarrow
\{1,2,3,4,5,6\}$ where $(x,y,z)$ are Cartesian coordinates. With this
notation, we have $c_{11} = S_{1111}$ and $c_{44} = S_{1212}$, while $c_{12} = S_{1122} +
P_\es$. (Note that Baiko's $C_{kl}$ denotes a different tensor.)

Since we are interested
solely in the lattice's response to shears, we can focus our
attention on a few components of this tensor. If we just consider the cases where a shear
strain yields a proportional shear stress (as is the case for isotropic materials), then
we will have contributions from the simple shear portions of the tensor, viz., $c_{44}$,
$c_{55}$, and $c_{66}$, in addition to the elongational shears $A_{12} := (c_{11} + c_{22})/2 - c_{12}$,
$A_{13} := (c_{11} + c_{33})/2 - c_{13}$, and $A_{23} := (c_{22} + c_{33})/2 - c_{23}$.
The simple shears correspond to stresses or strains of the form
$2\hat{x}_{(k}\hat{y}_{l)}$ where $\hat{x}_k$ and $\hat{y}_k$ are unit vectors in the $x$- and
$y$-directions, and parentheses on the indices indicate symmetrization.
The elongational shears correspond to
stresses or strains of the form $\hat{x}_{kl} - \hat{y}_{kl}$---i.e.,
rotations of the simple shears by $\pi/4$.
In the past, some investigations of shear modulus effects in the crusts of neutron
stars have used the simple shear portion of the tensor, $c_{44}$, as the shear
modulus of a bcc lattice (for which $c_{44} = c_{55} = c_{66}$)---see the
discussion in Sec.~7.1 of Chamel and Haensel~\cite{CH}. As
Strohmayer \emph{et al.}~\cite{Strohmayeretal} emphasize, it is inappropriate
to simply use one component of the elastic modulus tensor for this purpose.
While a truly detailed calculation would use the full elastic modulus tensor,
if one considers a polycrystal consisting of many randomly oriented domains,
it is possible to use an angle-averaged version of the shear portions of this
tensor and obtain an upper bound.

This upper bound is due to Voigt~\cite{Voigt}, and involves the elastic
constants given above---see Hill~\cite{Hill} for the proof that the
Voigt expression gives an upper bound.
We do not
consider any of the more involved sharper bounds, such as the oft-used
ones by Hashin and Shtrikman~\cite{HS}. (See~\cite{WDO} for a review of such
bounds, and~\cite{Watt} for a Hashin-Shtrikman bound for orthorhombic crystals,
such as the ones we consider here.)
The Voigt result was rediscovered by Ogata and Ichimaru~\cite{OI} in
considering the shear modulus of the neutron star crust. (They used
it as an average, as was originally proposed by Voigt, not as an upper bound,
as in Hill.)
Following \citet{OI} and most subsequent work, we shall use the Voigt average
for our effective shear modulus, giving
\<\label{mu_eff1}
\mu_\eff = (A_{12} + A_{13} + A_{23})/15 + (c_{44} + c_{55} + c_{66})/5.
\?
We note, as does \citet{Baiko}, that the lattice will likely tend to align
itself with the star's magnetic field, leading to large-scale ordered
structure.
However the standard averaging approach we use here is a necessary first step,
and may be a good approximation to bulk properties of the star if the internal
magnetic field lines are tangled as in some simulations such as
\citet{Braithwaite2004}.

To dimensionally continue this expression, we need to consider the effects of
different lattice dimensionalities on the shear elastic coefficients.
We shall first discuss the integer dimension cases,
and then present the dimensionally continued expression for the effective
shear modulus at the end. (See
Pethick
and Potekhin~\cite{PP} for related discussion about the elasticity of liquid
crystals as applied to the pasta phases in the crust.) Our discussion will be
simplified by the fact that we can take the elastic constants of all the
lattices to have cubic symmetry---this holds for the integer dimension lattices
and remains true for the dimensionally continued ones by fiat. However, there
will be differences between the elongational shear elastic constants in which
the shear takes place completely within the lattice and those that involve an
elongation along the rods or slabs. We also will find that the
number of nonzero shear elastic constants is reduced due to the translational
symmetry along the rods or slabs in lower dimensions. Specifically,
\begin{enumerate}
\item In the
three-dimensional case, all the shear elastic constants are nonzero, and, by
cubic symmetry, the simple and elongational shear elastic constants are each
all equal (i.e., $c_{44} = c_{55} = c_{66}$ and $A_{12} = A_{13} = A_{23} =: 
A_\lat$).
\item In the two-dimensional case, only one of the simple shear elastic
constants is nonzero, since simple shears along the rods do not lead to a shear
stress---i.e., we have $c_{44} \neq c_{55} = c_{66} = 0$, where we have taken
the rods to point in the $z$-direction. The elongational shears are all
nonzero, but are not all equal: In general, the one elongational shear perpendicular to
the rods (i.e., within the lattice, so denoted $A_\lat$) has a different elastic constant than do the two elongational shears
that involve elongations along the rods (i.e., perpendicular to the lattice, and so denoted $A_\perp$). Explicitly, we have
$A_\lat := A_{12} \neq A_{13} = A_{23} =: A_\perp$.
\item In the one-dimensional case, all of the simple shear elastic constants
are zero, due to the translational symmetry along the slabs, and the only
nonzero elongational shear constants are those that involve elongation along
the slabs. (We neglect any
change in energy due to shearing one of the slabs.)
We thus have $c_{44} = c_{55} = c_{66} = 0$ and, taking the $x$-direction to
be perpendicular to the slabs, $A_\perp := A_{12} = A_{13} \neq A_{23} = 0$.
\end{enumerate}

We can now obtain the dimensionally continued version by using the
dimensionally continued versions of the relevant basic combinatorial results:
The number of independent shears (either simple or elongational) completely within the lattice (i.e., perpendicular
to the rods and slabs in the integer dimension cases) is given by $d(d-1)/2$;
this gives the multiplicity of the contributions to $\mu_\eff$ from $c_{44}$
and $A_\lat$. The number of
independent elongational shears with one elongation perpendicular to the
lattice and one within it is given by $d(3-d)$; this gives the multiplicity of
the contributions from $A_\perp$. The dimensionally continued
version of Eq.~\eqref{mu_eff1} is thus
\<\label{mu_eff}
\mu_\eff = \frac{d}{15}\left[\frac{d-1}{2}A_\lat + (3-d)A_\perp\right] + \frac{d(d - 1)}{10}c_{44}.
\?

\section{Calculation of the elastic constants}
\label{elastic_calc}

Now looking at how to compute $A_\lat$, $A_\perp$, and $c_{44}$, we can follow
Fuchs~\cite{Fuchs} in expressing these quantities in terms of lattice sums.
Specifically, we write the electrostatic energy of an appropriately deformed
lattice as a lattice sum and takes derivatives with respect to the deformation to
obtain lattice sum expressions for the elastic constants. In $A_\perp$, we will
also need to consider the
contributions from the cell energy [obtained using Eq.~\eqref{cell_energy}].
One also expects there to be contributions to $A_\lat$ and $c_{44}$ due to
changing the shape of the unit cell~\cite{Zach, ChGl97}, but we neglect these in
our treatment.
We do not have to concern ourselves with such contributions for $A_\perp$,
even in principle, since we can take the compression inside the lattice to be
uniform in all directions.

We also follow
Fuchs in using the Ewald method to compute the resulting lattice sums.
(See, e.g.,~\cite{JR} for a modern exposition of the Ewald method. Additionally,
we make certain modifications to the method so that it is compatible with
dimensional continuation---these are discussed in Sec.~\ref{Ewald}.) In the
Ewald method, one computes the sum of a slowly convergent or divergent
series by expressing it as the
sum of two rapidly convergent series, introducing an Ewald screening
function and using the Poisson summation formula.
(Actually, the Poisson summation formula is not strictly applicable
to the functions being summed in many applications, which is what allows the
Ewald method to regularize divergent series, by much the same method as zeta function
regularization---see the Appendix for further discussion.)

The Fuchs expressions are obtained by computing the energy of the perturbed lattice. Explicitly, we write the electrostatic energy per unit cell of the lattice (i.e., the energy required to
remove a single blob) as
\<\label{W}
\begin{split}
W &= \frac{Q^2}{2}\Biggl[\sideset{}{'}\sum_{\vec{x}\in\Lambda}(\phi E)(\|\vec{x}\|) +
\frac{1}{\Omega}\sideset{}{'}\sum_{\vec{p}\in\Lambda^*}(\widehat{\phi E_c})(\|\vec{p}\|)\\
&\quad - \frac{(\widehat{\phi E})(0)}{\Omega}\Bigg].
\end{split}
\?
[Compare the expression for the potential in Eq.~(7) of Johnson and
Ranganathan~\cite{JR} and Fuchs's expression for the energy in Eq.~(10)
of~\cite{Fuchs}.]
Here $Q$ is the charge of a blob; $\Lambda$ is the lattice under consideration, with dual lattice $\Lambda^*$, and 
physical Voronoi cell volume $\Omega$.
We shall also use Conway and Sloane's~\cite{CS_lat} notation of
$\sqrt{\det\Lambda}$ for the Voronoi cell volume without the physical scaling. 
Primes on the sums denote the omission of the
zero vector; $E$ is the Ewald
screening function, with complement $E_\mathrm{c}(x) := 1 - E(x)$;
$\phi$ is the
potential due to one of the blobs [including the effects of (physical) charge screening]; and the circumflex denotes the Fourier transform of (here
three-dimensional) radial functions. (See Sec.~\ref{Ewald}
for our Fourier transform conventions.)

We now give the lattice sum expressions for the elastic constants. The most straightforward are
those for $c_{11}$ and $c_{44}$, which are of the form
$\left.(d^2W/d\epsilon^2)\right|_{\epsilon=0}/\Omega$, for an appropriate deformation parametrized
by $\epsilon$.
(Note that we need $c_{11}$ in order to compute $A_\perp$ and the
dimensionally continued $A_\lat$, even though that elastic constant does not appear in $\mu_\eff$
by itself.) The resulting expressions for elastic constants take the form
($\Fc\in\{c_{11},c_{44}\}$)
\begin{widetext}
\<\label{Fc}
\begin{split}
\Fc &= \frac{Q^2}{2\Omega}\Biggl\{\sideset{}{'}\sum_{\vec{x}\in\Lambda}\left[(\phi E)'(\|\vec{x}\|)\frac{\partial^2\|\vec{x}\|}{\partial\epsilon^2} + (\phi E)''(\|\vec{x}\|)\left(\frac{\partial\|\vec{x}\|}{\partial\epsilon}\right)^2\right]  + \frac{1}{\Omega}\sideset{}{'}\sum_{\vec{p}\in\Lambda^*}\Biggl[2\frac{(\widehat{\phi E_c})(\|\vec{p}\|)}{\Omega^2}\left(\frac{\partial\Omega}{\partial \epsilon}\right)^2\\
&\quad + (\widehat{\phi E_c})^\bullet(\|\vec{p}\|)\biggl(\frac{\partial^2\|\vec{p}\|^2}{\partial\epsilon^2} - \frac{2}{\Omega}\frac{\partial\Omega}{\partial \epsilon}\frac{\partial\|\vec{p}\|^2}{\partial\epsilon}\biggr) + (\widehat{\phi E_c})^{\bullet\bullet}(\|\vec{p}\|)\left(\frac{\partial\|\vec{p}\|^2}{\partial\epsilon}\right)^2\Biggr] -
\frac{2}{\Omega^3}\left(\frac{\partial\Omega}{\partial \epsilon}\right)^2(\widehat{\phi E})(0) \Biggr\}\Biggr|_{\epsilon=0},
\end{split}
\?
\end{widetext}
where the coordinate and $\Omega$ derivatives are given by Eqs.~\eqref{c44_derivs} and
\eqref{c11_derivs}. [Recall that $\Omega$ is unchanged by the $c_{44}$ perturbation.]
The superscript bullets ($\bullet$) denote derivatives taken with
respect to $\|\vec{p}\|^2$---cf.\ the discussion below Eq.~\eqref{c44_derivs}.

To obtain the derivatives needed for the lattice sum expression for $c_{44}$, we note
(following Fuchs~\cite{Fuchs}) that $c_{44}$ is the only elastic constant that
contributes to
the change of energy of the lattice if one considers a simple shear
deformation. If we take this simple shear to be in the $x_1$ and $x_2$ directions [now denoting
our Cartesian coordinates by $(x_1,x_2,x_3)$, instead of the previous $(x,y,z)$, for notational
convenience], it corresponds to the deformations  $x_1\to x_1 + \epsilon x_2$
for the direct lattice, and $p_2 \to p_2 - \epsilon p_1$ for the dual lattice (with all other coordinates held fixed).
The requisite derivatives are thus
\begin{subequations}\label{c44_derivs}
\begin{align}
\left.\frac{\partial\|\vec{x}\|}{\partial\epsilon}\right|_{\epsilon = 0} &= 
\frac{x_1x_2}{\|\vec{x}\|},&
\left.\frac{\partial^2\|\vec{x}\|}{\partial\epsilon^2}\right|_{\epsilon = 0} &=
\frac{x_2^2}{\|\vec{x}\|} - \frac{x_1^2x_2^2}{\|\vec{x}\|^3},\\
\left.\frac{\partial\|\vec{p}\|^2}{\partial\epsilon}\right|_{\epsilon = 0} &=
-2p_1p_2,&
\left.\frac{\partial^2\|\vec{p}\|^2}{\partial\epsilon^2}\right|_{\epsilon = 0} &=
2p_1^2.
\end{align}
\end{subequations}
[\emph{Nota bene} (N.B.): We have given the derivatives of $\|\vec{p}\|^2$ since the dimensionally continued
Fourier transform depends naturally on this quantity---cf.\ the second expression in Eq.~(6) in~\cite{NKJ-M}. Additionally, all of the
right-hand sides of these expressions are evaluated at $\epsilon = 0$ (both here and in all similar
situations in the next equation).]

For $c_{11}$, we use the deformation $x_1 \to (1 + \epsilon)x_1$, $p_1 \to p_1/(1+\epsilon)$
(with all other coordinates held fixed).
 This perturbation changes the cell volume, so we must include
derivatives of $\Omega$, giving
\begin{subequations}\label{c11_derivs} 
\begin{align}
\left.\frac{\partial\|\vec{x}\|}{\partial\epsilon}\right|_{\epsilon = 0} &=
\frac{x_1^2}{\|\vec{x}\|},&
\left.\frac{\partial^2\|\vec{x}\|}{\partial\epsilon^2}\right|_{\epsilon = 0} &=
\frac{x_1^2}{\|\vec{x}\|} - \frac{x_1^4}{\|\vec{x}\|^3},\\
\left.\frac{\partial\|\vec{p}\|^2}{\partial\epsilon}\right|_{\epsilon = 0} &=
-2p_1^2,&
\left.\frac{\partial^2\|\vec{p}\|^2}{\partial\epsilon^2}\right|_{\epsilon = 0} &=
6p_1^2,\\
\label{Omega_derivs}
\left.\frac{\partial\Omega}{\partial \epsilon}\right|_{\epsilon = 0} &= \Omega,&
\left.\frac{\partial^2\Omega}{\partial \epsilon^2}\right|_{\epsilon = 0} &= 0.
\end{align}
\end{subequations}

For $A_\lat$, Fuchs uses the elongational shear in the $x_1$ and $x_2$
directions, for which the change to the lattice's energy is given by $2A_\lat$
alone (and has a lattice sum expression similar to that for $c_{44}$). However, as we shall see in
Sec.~\ref{dim_cont}, the resulting
expression is not well suited for dimensional continuation, since it contains
fourth powers of more than one coordinate [see Eq.~(12) in
Fuchs~\cite{Fuchs}~]. We can obtain an
expression that \emph{is} well-suited to dimensional continuation if we note that
$A_\lat = c_{11} - c_{12}$ (we have $c_{11} = c_{22}$ by cubic symmetry): We compute
$c_{11}$ as above, but obtain $c_{12} = S_{1122} + P_\es$ by
first using a two-component
perturbation to obtain $S_{1122}$, and then computing $P_\es$ \emph{{\`{a}} la} Fuchs. The
two-component perturbation we use to obtain $S_{1122}$ is $x_1\to (1 + \epsilon_1)x_1$,
$x_2\to (1 + \epsilon_2)x_2$, $p_1\to p_1/(1 + \epsilon_1)$, $p_2\to p_2/(1 + \epsilon_2)$
(with $x_3$ and $p_3$ held fixed), yielding the derivatives
\begin{subequations}\label{S1122_derivs} 
\begin{align}
\left.\frac{\partial\|\vec{x}\|}{\partial\epsilon_j}\right|_{\epsilon_{1,2} = 0} &=
\frac{x_j^2}{\|\vec{x}\|},&
\left.\frac{\partial^2\|\vec{x}\|}{\partial\epsilon_1\partial\epsilon_2}\right|_{\epsilon
_{1,2} = 0} &=
-\frac{x_1^2x_2^2}{\|\vec{x}\|^3},\\
\left.\frac{\partial\|\vec{p}\|^2}{\partial\epsilon_j}\right|_{\epsilon_{1,2} = 0} &=
-2p_j^2,&
\left.\frac{\partial^2\|\vec{p}\|^2}{\partial\epsilon_1\partial\epsilon_2}\right|_{\epsilon_{1,2} = 0} &= 0,\\
\left.\frac{\partial\Omega}{\partial \epsilon_j}\right|_{\epsilon_{1,2} = 0} &=\Omega,&
\left.\frac{\partial^2\Omega}{\partial\epsilon_1\partial\epsilon_2}\right|_{\epsilon_{1,2} = 0} &= \Omega,
\end{align}
\end{subequations}
where $j\in\{1,2\}$ and $\epsilon_{1,2} = 0$ $\Rightarrow$ $\epsilon_1 = \epsilon_2 = 0$.
One then obtains, from $S_{1122} = \left.(\partial^2W/\partial\epsilon_1\partial\epsilon_2)\right|_{\epsilon_{1,2}=0}/\Omega$ and the expression for $W$ in Eq.~\eqref{W},
\<\label{S1122}
\begin{split}
S_{1122} &= \frac{Q^2}{2\Omega}\Biggl\{\sideset{}{'}\sum_{\vec{x}\in\Lambda}\frac{x_1^2x_2^2}{\|\vec{x}\|^2}\left[(\phi E)''(\|\vec{x}\|) - \frac{(\phi E)'(\|\vec{x}\|)}{\|\vec{x}\|}\right]\\
&\quad + \frac{1}{\Omega}\sideset{}{'}\sum_{\vec{p}\in\Lambda^*}\biggl[(\widehat{\phi E_c})(\|\vec{p}\|) + 2(p_1^2 + p_2^2)(\widehat{\phi E_c})^\bullet(\|\vec{p}\|)\\
&\quad + 4p_1^2p_2^2(\widehat{\phi E_c})^{\bullet\bullet}(\|\vec{p}\|)\biggr] - \frac{1}{\Omega}(\widehat{\phi E})(0)\Biggr\}\Biggr|_{\epsilon_{1,2}=0}.
\end{split}
\?
We calculate $P_\es$ using the same perturbation we used to obtain $c_{11}$ [with the derivatives
given in Eq.~\eqref{c11_derivs}]. Specifically,
$P_\es = -\left.(dW/d\epsilon)\right|_{\epsilon=0}/\Omega$ [cf.\ Eq.~\eqref{Sklps}, recalling that
$S_{kl} = -P_\es\delta_{kl}$]. This gives
\<\label{Pes}
\begin{split}
P_\es &= -\frac{Q^2}{2\Omega}\Biggl\{\sideset{}{'}\sum_{\vec{x}\in\Lambda}\frac{x_1^2}{\|\vec{x}\|}(\phi E)'(\|\vec{x}\|) + \frac{1}{\Omega}(\widehat{\phi E})(0)\\
&\quad - \frac{1}{\Omega}\sideset{}{'}\sum_{\vec{p}\in\Lambda^*}\biggl[(\widehat{\phi E_c})(\|\vec{p}\|) + 2p_1^2(\widehat{\phi E_c})^\bullet(\|\vec{p}\|)\biggr]\Biggr\}\Biggr|_{\epsilon=0}.
\end{split}
\?
We thus have
\<\label{Alat}
A_\lat = c_{11} - S_{1122} - P_\es,
\?
where $c_{11}$ is given by Eqs.~\eqref{Fc} and \eqref{c11_derivs}, and $S_{1122}$ and $P_\es$ are
given by the above expressions.

Now, to obtain $A_\perp$, we need to supplement the expression for $c_{11}$ with the contributions
due to changing the cell energy (since we are changing its radius), along with the contributions
due to changing the blobs' charge. (As discussed at the end of Sec.~4 of Baiko~\cite{Baiko}, one does not
expect these to contribute to $A_\lat$, and, indeed, one can show that the individual contributions all cancel.  They make a nonzero contribution to $A_\perp$ due to anisotropy.) Additionally, we shall see that there is a
contribution from $P_\es$ due to the stressed background. The contribution due to changing the cell energy is
\<\label{Apcell}
A_{\perp,\cell} = \frac{2}{d^2}\frac{E_\cell}{\Omega},
\?
where  $E_\cell/\Omega$ is given in Eq.~\eqref{cell_energy}.
This can be deduced from the scalings of the Coulomb and surface contributions [see Eq.~(2) in
Pethick and Potekhin~\cite{PP}~], noting that the change in the cell radius with this perturbation
is given by
\<
\left.\frac{\partial r}{\partial \epsilon}\right|_{\epsilon = 0} = \frac{r}{d}, 
\?
which comes from noting that $\Omega = C_\cell r^d$, where $C_\cell$ is a constant (since $x$ is
fixed, as we are keeping the overall density fixed) and
$(\partial\Omega/\partial\epsilon)|_{\epsilon = 0} = \Omega$ for the $c_{11}$ perturbation.
For the derivatives of the blobs' charge, we note that the (three-dimensional)
charge
density is fixed, so that the derivatives of the blobs' charge can be obtained from those of the cell
volume [in Eq.~\eqref{Omega_derivs}] by replacing the cell volume with the blobs' charge. Explicitly,
we have
\<
\left.\frac{\partial Q}{\partial \epsilon}\right|_{\epsilon = 0} = Q,\qquad
\left.\frac{\partial^2 Q}{\partial \epsilon^2}\right|_{\epsilon = 0} = 0.
\?

We now show how to put together all these contributions to
obtain the sum that gives $A_\perp$. First, we note that (with either $d = 2$ or $d = 1$)
\<\label{Aperp_orig}
\begin{split}
A_\perp &= \frac{1}{2}B_{klps}(\hat{x}_{kl} - \hat{z}_{kl})(\hat{x}_{ps} - \hat{z}_{ps})\\
&= \frac{1}{2}\left(S_{1111} + S_{3333} - S_{1133} - S_{3311} - 2P_\iso - P_\es\right),
\end{split}
\?
while we have
\<
\begin{split}
W &= W_0 + \{\text{terms linear in }\epsilon\} - P_\iso\epsilon^2\\
&\quad + \frac{1}{2}\epsilon^2(S_{1111} + S_{3333}  - S_{1133} - S_{3311}) + O(\epsilon^3),
\end{split}
\?
with the $x\to (1 + \epsilon)x$, $z \to z/(1+\epsilon)$ perturbation, so
\<
\frac{1}{\Omega}\left.\frac{d^2W}{d\epsilon^2}\right|_{\epsilon=0} = S_{1111} + S_{3333}  - S_{1133} - S_{3311} - 2P_\iso.
\?
Thus, since this perturbation corresponds to a $c_{11}$ perturbation within the lattice with a
corresponding elongation orthogonal to the lattice, we have
\<
S_{1111} + S_{3333}  - S_{1133} - S_{3311} - 2P_\iso = c_{11} + A_{\perp, \cell} + A_{\perp,Q}.
\?
Combining this together with Eq.~\eqref{Aperp_orig}, we thus obtain
\<\label{Aperp}
A_\perp =  \frac{1}{2}\left(c_{11} + A_{\perp,Q} + A_{\perp,\cell} - P_\es\right),
\?
where $c_{11}$ is given by Eqs.~\eqref{Fc} and \eqref{c11_derivs},
$A_{\perp,\cell}$ is given in Eq.~\eqref{Apcell}, $P_\es$ is given in Eq.~\eqref{Pes}, and
\<\label{ApQ}
\begin{split}
A_{\perp,Q} &= \frac{Q^2}{\Omega}\Biggl\{\sideset{}{'}\sum_{\vec{x}\in\Lambda}\left[(\phi E)(\|\vec{x}\|) + 2\frac{x_1^2}{\|\vec{x}\|}(\phi E)'(\|\vec{x}\|)\right]\\
&\quad  - \frac{1}{\Omega}\sideset{}{'}\sum_{\vec{p}\in\Lambda^*}\left[(\widehat{\phi E_c})(\|\vec{p}\|)
 + 4p_1^2(\widehat{\phi E_c})^\bullet(\|\vec{p}\|)\right]\\
&\quad + \frac{(\widehat{\phi E})(0)}{\Omega}\Biggr\}\Biggr|_{\epsilon=0}.
\end{split}
\?
As one would expect from Earnshaw's theorem, 
$c_{11}$ is always negative, so one relies on the contributions from
$A_{\perp,\cell}$, $A_{\perp,Q}$, and $-P_\es$ ($P_\es$ is negative) to make $A_\perp$ positive so that the
lattice is stable to shears. (See
Sec.~\ref{results1} for further discussion.)

\section{Dimensional continuation of lattice sums}
\label{dim_cont}

\subsection{The dimensionally continued Ewald method}
\label{Ewald}

In order to compute these lattice sums numerically, we employ a
generalization of the standard Ewald~\cite{Ewald} method used
by Fuchs~\cite{Fuchs}.
Showing the
standard integer
dimension version first, we have, summing a function $f$ over a
lattice $\Lambda$ with dual lattice $\Lambda^*$,
\<
\sum_{\vec{x} \in \Lambda} f(\vec{x}) = \sum_{\vec{x} \in \Lambda} (fE)(\vec{x}) + \frac{1}{\sqrt{\det \Lambda}}\sum_{\vec{p} \in \Lambda^*}(\widetilde{fE_\mathrm{c}})(\vec{p}).
\?
Here $E_\mathrm{c}(\vec{x}) := 1 - E(\vec{x})$ is the complement of the
Ewald screening function $E$, $\tilde{g}(\vec{p}) :=
\int_{\R^n} g(\vec{x})e^{-2\pi i\vec{x}\cdot\vec{p}}d^nx$
denotes the standard Fourier transform, and we choose $E$ so that both of the
sums on the right-hand side converge quickly. [Recall that the classical
Poisson summation formula says that $\sum_{\vec{x}\in\Lambda} f(\vec{x}) =
(\det\Lambda)^{-1/2}\sum_{\vec{p}\in\Lambda^*}\tilde{f}(\vec{p})$.]

The classic choice for $E$ for Coulombic potentials 
(dating back to Ewald)
is the complementary error function. However, this turns out to be insufficiently
flexible to provide good convergence for the sums we consider (particularly for
small $d$). Following Nijboer and de Wette~\cite{NdW} and
Fortuin~\cite{Fortuin}, we introduce the incomplete gamma function,
$\Gamma(\cdot,\cdot)$, and use the screening function
\<
E(\vec{x}) = \Gamma(N/2,\alpha^2\|\vec{x}\|^2)/\Gamma(N/2).
\?
This
reduces to Ewald's complementary error function for $N = 1$. The extra freedom contained in
$N$ allows us to tune $E$ to provide fast convergence for the sums we encounter. We used $N = 10$ and
$\alpha = 1.2/a$ in the computations reported in
Sec.~\ref{results1}. [Here $a$ is the lattice spacing, given in Eq.~\eqref{scaling}.]
We could have doubtless obtained faster convergence if we
had allowed these parameters to vary with $d$ (and possibly also $\lambda$),
particularly for $d$ close to $1$. However, we found these values to give
reasonably good performance, and thus did not perform much experimentation
beyond checking that our results are insensitive to small variations in the
Ewald screening parameters. 

In order to dimensionally continue our lattice sums, we need to dimensionally
continue the Poisson summation formula. We shall give an overview of the
calculational aspects here---see~\cite{NKJ-M} for more details of the derivations, and a proof
of the formula for nicely behaved functions.
We first introduce the dimensionally continued Fourier transform for spherically
symmetric functions, given in Theorem~3.3 of Chap.~IV of Stein and
Weiss~\cite{SW},
\<\label{g-hat}
\hat{g}(p) := \frac{2\pi}{p^{d/2-1}}\int_0^\infty g(r)J_{d/2-1}(2\pi p
r)r^{d/2}dr.
\?
Here $J_\nu$ is the Bessel function of the first kind.
(See Sec.~2.2 of~\cite{NKJ-M} for more details, including an alternative
expression in terms of a hypergeometric function.)
We also note that we can compute the dimensionally continued Fourier transform
of the potential from its defining partial differential equation~\eqref{phiPDE}, giving
\<\label{phi-hat}
\hat{\phi}(p) = \frac{4\pi}{4\pi^2p^2 + \lambda^{-2}}.
\?
[This agrees with the result of the more involved calculation one could perform
using the dimensionally continued Fourier integral given in Eq.~\eqref{g-hat}.]
We use Eq.~\eqref{phi-hat} along with the dimensionally continued Fourier
integral~\eqref{g-hat} to compute $\widehat{\phi E_c}$ efficiently,
writing it as $\hat{\phi} - \widehat{\phi E}$, where the integral giving the
second term converges reasonably rapidly.

We then introduce the theta series of a lattice.
As is discussed in more detail in Sec.~2.3 of Chap.~2 of Conway and Sloane~\cite{CS_lat},
the theta series of a lattice is the generating function of the number of lattice
points on a sphere of a given squared radius, so the theta series of a lattice
$\Lambda$ is defined by
\<\label{ThetaLambda}
\Theta_\Lambda(q) := \sum_{\vec{k}\in\Lambda}q^{\|\vec{k}\|^2}.
\?
[N.B.: There are several different notational conventions for theta series and theta functions. We
have chosen to write all our theta series and functions as functions of the nome, $q$, unless we note
otherwise. These exceptions will only occur in discussions of the Jacobi formula for the theta series of the dual lattice, where it is
convenient to treat the theta series as a function of a complex variable $z$, with $q = e^{i\pi z}$.
We shall denote this by an overbar---e.g., $\bar{\Theta}_{\Lambda}(z) := \Theta_\Lambda(e^{i\pi z})$. Conway and Sloane treat all their theta functions as
functions of $z$, even when they write their expansions in terms of the nome.]
Thus, if we define the power series coefficients of the theta series using
\<
\Theta_\Lambda(q) =: \sum_{l=0}^\infty A_l q^{B_l},
\?
we can write the sum of a spherically symmetric function $F$ over
$\Lambda$ as
\<
\sum_{\vec{k}\in\Lambda}F(\|\vec{k}\|) = \sum_{l=0}^\infty A_l F(\sqrt{B_l}).
\?

The dimensionally continued Poisson summation formula then has the form
\<
\sum_{l=0}^\infty A_l F(\sqrt{B_l}) =
\frac{1}{\sqrt{\det\Lambda}}\sum_{l=0}^\infty A_l^*\hat{F}(\sqrt{B_l^*}).
\?
Here, $A_l^*$ and $B_l^*$ are defined
analogously to their counterparts for $\Lambda$. The theta series for
$\Lambda^*$ can be
calculated from $\Theta_\Lambda$ using the Jacobi formula~\eqref{JTF}, which is already
in dimensionally continued form. [N.B.: The dimensionally continued Poisson summation formula
presented in~\cite{NKJ-M} omits the factor of $(\det\Lambda)^{-1/2}$, since, as
discussed there, this factor cancels against a similar one present in the Jacobi transformation
formula. We include the factor here, since we obtain a simpler result for the theta series of the dual lattice
by using the standard Jacobi transformation formula.]
Of course, we need to sum more than just spherically symmetric functions
[see Eqs.~\eqref{Fc}--\eqref{Pes}],
but, as we shall see shortly, this dimensionally
continued Poisson summation formula will be sufficient for our needs.

\subsection{Dimensional continuation of the lattice}
\label{lat_dim_cont}

Turning now to the problem of dimensionally continuing the lattices themselves,
recall that the integer dimension lattices are [up to an overall scaling, which we 
determine in Eq.~\eqref{scaling}]
a bcc lattice
for $d = 3$, a hexagonal lattice for $d = 2$, and $\Z$,
the (one-dimensional) lattice of integers,
for $d = 1$. In some sense, the natural way to dimensionally continue these
lattices would be to dimensionally continue the root lattice family $A_d^*$,
which gives those lattices for $d\in\{1,2,3\}$ and, more generally, gives the
best lattice covering of $\R^n$ for $n\leq 5$ (as discussed in Conway and
Sloane~\cite{CS_lat}). However, unlike most other families of lattices, $A_d^*$
does not
have a theta series that is written in a nicely dimensionally continued form.
Its theta series is written in terms of a sum whose
number of terms depends upon dimension---see, e.g., Eq.~(56) in Chap.~4 of Conway and
Sloane~\cite{CS_lat} for the theta series for $A_d$, from which the theta series for the dual lattice
can be obtained using Jacobi's formula [our Eq.~\eqref{JTF}]. The sum is over $d$th roots of
unity, so one could contemplate writing it as an integral, using Cauchy's theorem, and proceeding
that way. However, even disregarding the complications this would involve, it is not clear how to
compute the sums involving $x_1^4$ that we need (for, e.g., $c_{11}$) in this
framework, or, alternatively, how to
implement the requisite distortions to the lattice at the level of its theta series.

We thus proceed by treating the dimensionally
continued lattice as a union of hyperlattices (i.e., lattices
of one dimension fewer than the overall lattice)\footnote{Note that many of
these hyperlattices are actually shifted lattices, mathematically
speaking.} whose separation is given
by a freely specifiable function $f^\lat$ that interpolates between the integer
dimension separations.  One then finds that the sums over the hyperlattices
dimensionally continue in a natural way, and that the final result for the shear modulus is
rather insensitive to the choice of $f^\lat$ (provided that it satisfies some reasonable properties,
discussed below).

\begin{figure}[htb]
\begin{center}
\epsfig{file=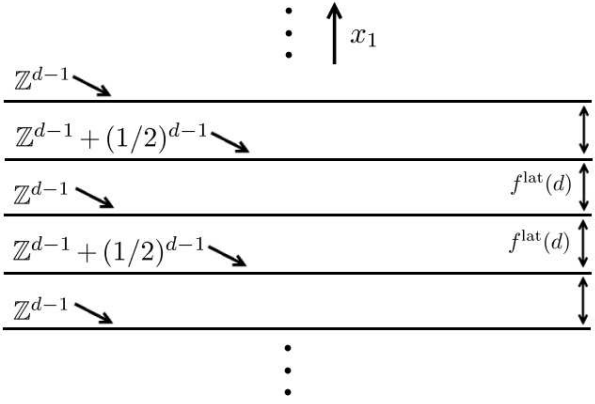,width=8cm,clip=true}
\end{center}
\caption[Decomposition of the full lattice into hyperlattices]{\label{lattice_fig} A schematic of the decomposition of the full
lattice into hyperlattices.
}
\end{figure}

Explicitly, the (scaled) integer dimension lattices can be written as
$[f^\lat(d)\Z_\even]\times \Z^{d-1} + [f^\lat(d)\Z_\odd]\times[\Z^{d-1}+(1/2)^{d-1}]$,
where $\Z_\even$ and $Z_\odd$ denote the even and odd integers respectively,
$\Z^{d-1}$ denotes the $(d-1)$-dimensional lattice of integers, and
$\Z^{d-1}+(1/2)^{d-1}$ denotes the same shifted by the [$(d-1)$-dimensional]
vector all of whose components are $1/2$.
This is illustrated in Fig.~\ref{lattice_fig}, where we take $x_1$ to be in
the direction orthogonal to the hyperlattices.

\begin{figure}[htb]
\begin{center}
\epsfig{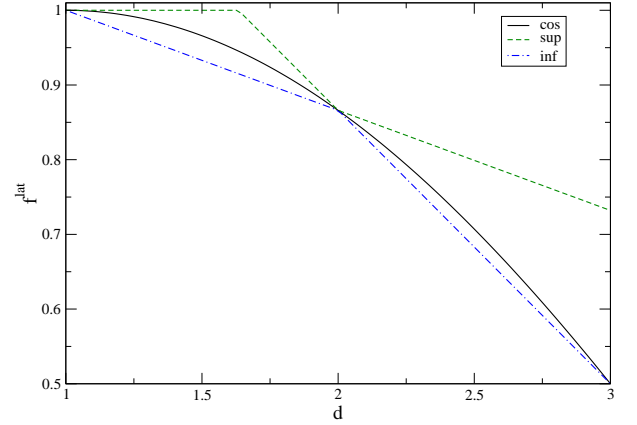}
\end{center}
\caption[Lattice interpolation functions]{\label{f_comp} Plots of $f^\lat_\mathrm{cos}$ [given in
Eq.~\eqref{f_lat}], along with the pointwise sup and inf over all such
possibilities [$f^\lat_\mathrm{sup}$ and $f^\lat_\mathrm{inf}$, given in
Eqs.~\eqref{f_bounds}]. 
}
\end{figure}

Here $f^\lat(d)$ is freely specifiable, except that it must satisfy
$\{1,2,3\} \mapsto \{1, \sqrt{3}/2, 1/2\}$. Additionally, it makes sense to choose
it to be smooth, nonincreasing, and concave.
A possibility that satisfies all these criteria is
\<\label{f_lat}
f^\lat_\mathrm{cos}(d) := \cos\left(\frac{\pi}{6}[d-1]\right),
\?
which we will use in all our results, unless otherwise noted.
We will show that the results for other possibilities all agree well with
$f^\lat_\mathrm{cos}$.
Consider the envelope of all possible functions satisfying the requirements.
The bounds on this envelope are
\begin{subequations}
\label{f_bounds}
\begin{align}
f^\lat_\mathrm{inf}(d) &=
\begin{cases}
(\sqrt{3}/2 - 1)(d-1) + 1, & 1\leq d < 2,\\
[(1 - \sqrt{3})(d-2) + \sqrt{3}]/2, & 2\leq d\leq 3,
\end{cases}\\
f^\lat_\mathrm{sup}(d) &=
\begin{cases}
1, & 1 \leq d < d',\\
[(1 - \sqrt{3})(d-2) + \sqrt{3}]/2, & d'\leq d < 2,\\
(\sqrt{3}/2 - 1)(d-1) + 1, & 2\leq d\leq 3,
\end{cases}
\end{align}
\end{subequations}
where $d' := 3 - 1/(\sqrt{3}-1) \simeq 1.63$. See Fig.~\ref{f_comp} for an
illustration. While it would be natural to choose $f^\lat$ so that the dimensionally
continued lattice had effective cubic symmetry (discussed in Sec.~\ref{lattice_sums}), it is not possible to do so.
If we consider, for instance, $d = 5/2$, then the $\Z^{d-1}$ hyperlattices have theta series $1 + 3q + O(q^2)$,
while the theta series for the $\Z^{d-1} + (1/2)^{d-1}$ hyperlattices is $2^{3/2}q^{3/8}[1+ O(q^2)]$.
The irrational prefactor in the second theta series is a clear obstruction to obtaining effective
cubic symmetry: There is no way of obtaining an irrational contribution from the first theta
series, since all its coefficients are rational.

Now, we dimensionally continue the sums over the hyperlattices using the
hyperlattice's theta series, which are naturally dimensionally continued.
Specifically, the
theta series of $\Z^d$ and $\Z^d + (1/2)^d$ are
$\vartheta_3^d$ and $\vartheta_2^d$, where
\begin{subequations}
\begin{gather}
\label{thetas}
\vartheta_2(q) := \sum_{k\in\Z}q^{(k+1/2)^2},\qquad \vartheta_3(q) := \sum_{k\in\Z}q^{k^2},\\
\vartheta_4(q) := \sum_{k\in\Z}(-q)^{k^2}.
\end{gather}
\end{subequations}
See, e.g., Sec.~5 of Chap.~4 of Conway and Sloane~\cite{CS_lat}, but
recall the differences in their theta function notation, discussed below
Eq.~\eqref{ThetaLambda}.
We also introduce
$\vartheta_4$ here since it appears in the theta series of the dual
lattice.

The theta series of the full lattice is thus
\<\label{theta_Lambda}
\begin{split}
\bar{\Theta}_{\Lambda}(z) &= \sum_{k \in \Z}\biggl\{q^{[2kf^\lat(d)]^2}\left[\bar{\vartheta}_3(z)\right]^{d-1}\\
&\quad + q^{[(2k + 1)f^\lat(d)]^2}\left[\bar{\vartheta}_2(z)\right]^{d-1}\biggr\}\\
&= \bar{\vartheta}_3([2f^\lat(d)]^2z)\left[\bar{\vartheta}_3(z)\right]^{d-1}\\
&\quad +
\bar{\vartheta}_2([2f^\lat(d)]^2z)\left[\bar{\vartheta}_2(z)\right]^{d-1}.
\end{split}
\?
One can check that this reduces to the appropriate expressions for
$d\in\{1,2,3\}$; the theta series for the $3$-dimensional bcc lattice and
$2$-dimensional hexagonal lattice are given in Eqs.~(96) and~(60)
of Chap.~4 of Conway and Sloane~\cite{CS_lat}. For $d = 1$, we obtain the theta
series for $\Z$ in a nonstandard form---one can convert it to the standard one
given above by using the identity $\bar{\vartheta}_2(4z) +
\bar{\vartheta}_3(4z) = \bar{\vartheta}_3(z)$ from Eq.~(22) in Chap.~4 of
Conway and Sloane.

We use Jacobi's formula (which is already dimensionally continued) to obtain the theta series of the
dual lattice, viz., [e.g., Eq.~(4) in~\cite{NKJ-M}, or Eq.~(19) in Chap.~4 of Conway and
Sloane~\cite{CS_lat}~]
\<\label{JTF}
\bar{\Theta}_{\Lambda^*}(z) = \sqrt{\det\Lambda}(i/z)^{d/2}\bar{\Theta}_\Lambda(-1/z).
\? 
When applying Jacobi's formula, we take the volume of the lattice's Voronoi cell to be
dimensionally continued in the obvious way, viz., $\sqrt{\det\Lambda} = f^\lat(d)$. [N.B.: This expression neglects the overall scaling of the lattice, which
we fix in Eq.~\eqref{scaling}.]
We then obtain, upon use
of the theta function identities in Eq.~(21) of Chap.~4 in Conway and Sloane,
\<
\begin{split}
\bar{\Theta}_{\Lambda^*}(z) &=
\bigl\{\bar{\vartheta}_3(z/[2f^\lat(d)]^2)\left[\bar{\vartheta}_3(z)\right]^{d-1}\\
&\quad +
\bar{\vartheta}_4(z/[2f^\lat(d)]^2)\left[\bar{\vartheta}_4(z)\right]^{d-1}\bigr\}/2\\
&= \frac{1}{2}\sum_{k \in \Z}\biggl\{q^{[k/f^\lat(d)]^2}\left(\left[\bar{\vartheta}_3(z)\right]^{d-1} + \left[\bar{\vartheta}_4(z)\right]^{d-1}\right)\\
&\quad + q^{[(k + 1/2)/f^\lat(d)]^2}\left(\left[\bar{\vartheta}_3(z)\right]^{d-1} - \left[\bar{\vartheta}_4(z)\right]^{d-1}\right)\biggr\}.
\end{split}
\?
The second expression tells us that we can treat the dual lattice as a union of
hyperlattices separated by a distance of $[2f^\lat(d)]^{-1}$, where the even hyperlattices are
$D_{d-1}$ and the odd ones are $D_{d-1} + (0^{d-2} 1)$ (cf.\ the schematic of
the direct lattice shown in Fig.~\ref{lattice_fig}). Here $D_d$ is the
$d$-dimensional root lattice discussed in Sec.~7.1 of Chap.~4 of Conway and
Sloane~\cite{CS_lat}, and $D_d + (0^{d-1} 1)$ denotes the same lattice shifted by
a unit in one coordinate direction. These lattices' theta series are 
[Eqs.~(87) and (89) of Chap.~4 of Conway and Sloane~\cite{CS_lat}~]
$(\vartheta_3^d \pm \vartheta_4^d)/2$, with the upper [resp.\ lower] sign
corresponding to $D_d$ [resp.\ $D_d + (0^{d-1} 1)$].
This
interpretation is convenient,
as it allows us to compute the sums over the dual lattice using the same technology
as for the direct lattice.

\subsection{Lattice sums}
\label{lattice_sums}

From the expressions in Eqs.~\eqref{Fc}--\eqref{Pes}, we see that we need to dimensionally continue sums
involving a spherically symmetric function times either
$x_1^n$ ($n\in\{0,2,4\}$), $x_2^2$, or $x_{1}^2x_2^2$. In order to do this with our method, we note that we can express $x_1^n$ in terms of $f^\lat(d)$ and the index of
the hyperlattice ($k$, in our discussion below), due to our method of dimensionally continuing the lattice.
For $x_2^2$, we note that the hyperlattices have effective cubic symmetry, so we have
\<
\sum_{\vec{x}\in\cH_{d-1}\cap S_r}x_2^2 = \frac{n_\cH(r)r^2}{d-1},
\?
where
$\cH_{d-1}\cap S_r$ 
denotes the intersection of a
hyperlattice $\cH_{d-1}$ (with dimension $d-1$) with a $(d-2)$-sphere of radius $r$ centered at
the origin, and $n_\cH(r)$ is the number of points in this intersection
(obtained from $\cH_{d-1}$'s theta series). Unfortunately, this method is not applicable to
$x_2^4$ (which appears in the Fuchs expression for $A_\lat$ along with $x_1^4$), which is why
we compute $A_\lat$ using the more involved method discussed above. However, note that we can
compute $P_\es$ (and thus $A_\lat$) and $c_{44}$ two ways---the lattice sums we use do not contain 
fourth powers of any of the coordinates, so we can also compute them with the index substitution $1
\leftrightarrow 2$. We find that the two choices differ by $\lesssim10\%$. Since these would be
exactly equal if effective cubic symmetry held, this gives an indication
that our assumption of such symmetry is valid to about this level.

We now give the explicit expressions for the computation of the requisite lattice sums in a dimensionally
continued manner. If we consider
summing $G(x_1,x_2)F(\|\vec{x}\|)$ over the $d$-dimensional lattice $\Lambda_d$, where $G(x_1,x_2)$
is either one of $x_1^n$ ($n\in\{0,2,4\}$), $x_2^2$, or $x_1^2x_2^2$, then we have to evaluate the
following double sum:
\<
\begin{split}
\label{direct-sum}
\sum_{\vec{x} \in \Lambda_d}G(x_1,x_2)F(\|\vec{x}\|) &=
\sum_{k=0}^\infty\sum_{l=0}^\infty \Bigl\{\iota_k N^{[3]}_l\cA_{G} F\bigl(\cR^{[3]}\bigr)\\
&\quad + 2  N^{[2]}_l\cB_{G} F\bigl(\cR^{[2]}\bigr)\Bigr\},
\end{split}
\?
where
\begin{subequations}
\begin{align}
\cA_{x_1^n} &= [2k f^\lat(d)]^n, \quad &\cB_{x_1^n} &= [(2k + 1)f^\lat(d)]^n,\\
\cA_{x_2^2} &= l/(d-1), \quad &\cB_{x_2^2} &= l/(d-1) + 1/4,
\end{align}
\begin{align}
\cR^{[3]} &:= \sqrt{[2k f^\lat(d)]^2 + l},\\
\cR^{[2]} &:= \sqrt{[(2k + 1)f^\lat(d)]^2 + l + (d - 1)/4},
\end{align}
\end{subequations}
and one obtains the coefficients for the $x_1^2x_2^2$ case by multiplication
(e.g., $\cA_{x^2y^2} = \cA_{x^2}\cA_{y^2}$). [The
superscripts $[2]$ and $[3]$ come from the names of
the theta functions that give the pertinent hyperlattices' theta series---see
Eq.~\eqref{Ns}.]
The $l$-sum comes from summing over an individual hyperlattice, and the $k$-sum then sums over all
hyperlattices. We have introduced
\<
\iota_k :=
\begin{cases}
1,& k=0,\\
2,& \mbox{otherwise},
\end{cases}
\?
so we can take $k$ to run only over the positive integers,
while the hyperlattice index runs over \emph{all} integers.
The structure of the hyperlattices is accounted for by the theta series coefficients $N^{[j]}_l$, 
defined by
\<\label{Ns}
\sum_{l=0}^\infty N^{[3]}_lq^{l} := [\vartheta_3(q)]^{d-1},\quad
\sum_{l=0}^\infty N^{[2]}_lq^{l} := \left[\frac{\vartheta_2(q)}{q^{1/4}}\right]^{d-1}
\?
[recall that the hyperlattices' theta series are $\vartheta_2^{d-1}$ and
$\vartheta_3^{d-1}$; the theta functions are defined in Eq.~\eqref{thetas}].
Similarly, for the dual lattice, we have
\<
\begin{split}
\label{dual-sum}
\sum_{\vec{p}\in\Lambda_d^*}G(p_1,p_2)F(\|\vec{p}\|) &=
\sum_{k=0}^\infty\sum_{l=0}^\infty \Bigl\{\iota_k N^{+}_l\cC_{G} F\bigl(\cR^+\bigr)\\
&\quad + 2  N^{-}_l \cD_{G}F\bigl(\cR^-\bigr)\Bigr\},
\end{split}
\?
where
\begin{subequations}
\begin{align}
\cC_{p_1^n} &= [k/f^\lat(d)]^n, \quad &\cD_{p_1^n} &= [(k+1/2)/f^\lat(d)]^n,\\
\cC_{p_2^2} &= 2l/(d-1), \quad &\cD_{p_2^2} &= (2l+1)/(d-1),
\end{align}
\begin{align}
\cR^+ &:= \sqrt{[k/f^\lat(d)]^2 + 2l},\\
\cR^- &:= \sqrt{[(k + 1/2)/f^\lat(d)]^2 + 2l + 1},
\end{align}
\end{subequations}
(with the same multiplication for the $p_1^2p_2^2$ case as for the direct lattice),
and the theta series coefficients are given by
\<\label{Npm}
\sum_{l=0}^\infty N^{\pm}_lq^{l} := \left\{[\vartheta_3(q)]^{d-1} \pm [\vartheta_4(q)]^{d-1}\right\}/2
\?
[recall that the dual hyperlattices' theta series are $\vartheta_3^{d-1} \pm
\vartheta_4^{d-1}$].

We can now calculate the elastic constants by combining together these results
with our previously derived expressions. For $c_{44}$,
these are given by Eqs.~\eqref{Fc} and \eqref{c44_derivs}. For $A_\lat$ and $A_\perp$, one uses
the sums given in Eqs.~\eqref{Alat} and~\eqref{Aperp}, respectively.
We also need to determine the scaling of the lattice for a given energy
density. This is given by the cell radius,
$R$, computed in
Sec.~\ref{lat_struc}: Equating the volume of a cell with this radius to the volume of the
lattice's Voronoi cell [given by $f^\lat(d)a^d$] determines the lattice's spacing, $a$. Explicitly, we
have
\<\label{scaling}
a = \frac{\pi^{1/2}}{[f^\lat(d)\Gamma(d/2+1)]^{1/d}}R,
\?
using the dimensionally continued expression for the volume of a unit $d$-ball,
viz., $\pi^{d/2}/\Gamma(d/2+1)$.
Additionally, we compute $Q$, the charge of the blob, using the charge density of the rare
phase 
and the ($d$-dimensional) volume of the blob. 

\section{Results and discussion}
\label{results1}

\begin{figure}[htb]
\begin{center}
\epsfig{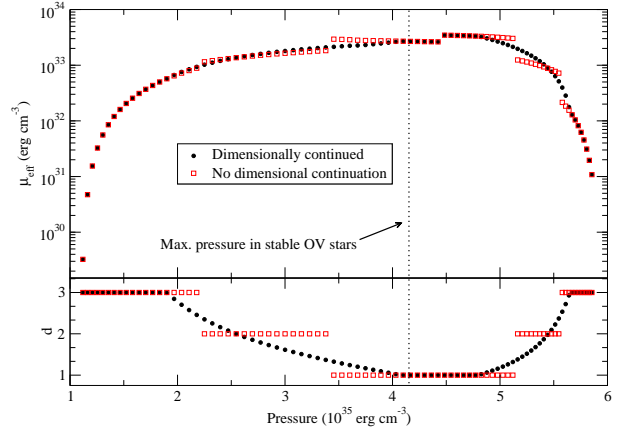}
\end{center}
\caption{\label{mu_eff_d_vs_p_Hy1_sigma80} The effective shear modulus and
lattice dimensionality versus pressure for the Hy1 EOS both with and without
dimensional continuation. We have also indicated the maximum pressure one obtains in
a stable OV star using this EOS.
}
\end{figure}

\begin{figure}[htb]
\begin{center}
\epsfig{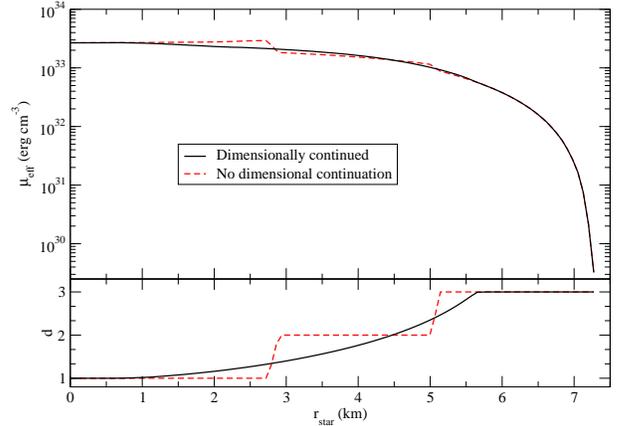}
\end{center}
\caption{\label{mu_eff_d_vs_r_Hy1_sigma80} The effective shear
modulus and
lattice dimensionality versus OV (Schwarzschild coordinate) radius for the 
maximum mass stable star (of total radius $12.5$~km) with the Hy1
EOS both with and without
dimensional continuation.
}
\end{figure}

Here we present the shear moduli for the various EOS parameters we
consider (given in Table~\ref{EOS_table}).

First, to give an indication of the effects of dimensional continuation, in
Fig.~\ref{mu_eff_d_vs_p_Hy1_sigma80} we
plot the effective shear modulus $\mu_\eff$ [see Eq.~\eqref{mu_eff}] and
lattice dimensionality $d$ versus pressure with and without dimensional
continuation, for the Hy1 EOS with a
surface tension of
$\sigma = 80 \text{ MeV fm}^{-2}$ [and the $f^\lat_\mathrm{cos}$ lattice
interpolation function from Eq.~\eqref{f_lat}]. We have shown all the pasta phases, even
though only the first few appear in stable stars, as is indicated in the
figure. The jump in the effective shear modulus occurs at the halfway
point---i.e., equal amounts of quark and hadronic matter---and is due to the
switch on the screening length $\lambda$ at that point (see the discussion in
Sec.~\ref{screening}). Also note that the lattice becomes unstable for a small
range of $d$ slightly less than $3$---see the discussion below. We additionally
show $\mu_\eff$ and $d$ versus the (Schwarzschild coordinate) OV stellar radius
$r_\mathrm{star}$ for the maximum mass star in
Fig.~\ref{mu_eff_d_vs_r_Hy1_sigma80}.

\begin{figure}[htb]
\begin{center}
\epsfig{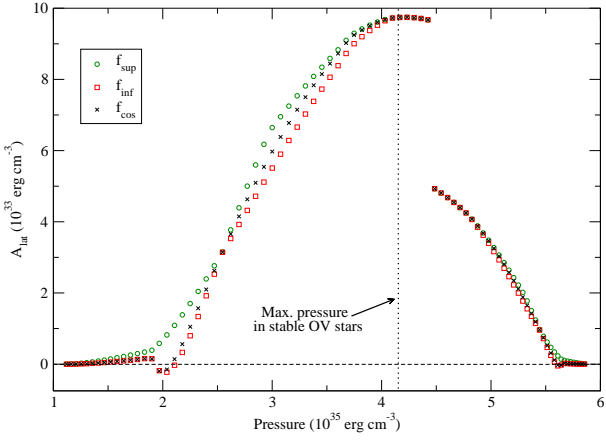}
\end{center}
\caption[$A_\lat$
vs.\ pressure for the Hy1 EOS and a variety of surface
tensions]{\label{Alat_vs_p_Hy1_f_lats} The (dimensionally continued) $A_\lat$
versus pressure for the Hy1 EOS and a variety of $f^\lat$s.
See the text for a discussion of the small unphysical dip below zero.
}
\end{figure}

\begin{figure}[htb]
\begin{center}
\epsfig{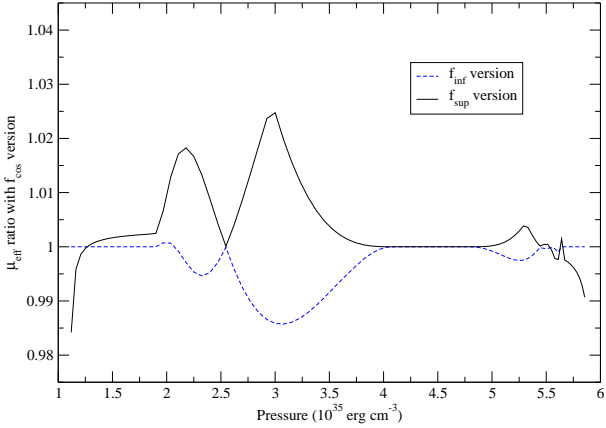}
\end{center}
\caption[Ratios of the effective shear modulus using different lattice
interpolation functions]{\label{mu_eff_f_lat_ratios} Ratios of the
effective shear modulus computed using $f^\lat_\mathrm{sup}$ and
$f^\lat_\mathrm{inf}$ to that computed using $f^\lat_\mathrm{cos}$ [see
Eqs.~\eqref{f_lat} and~\eqref{f_bounds}] plotted versus pressure for the Hy1 EOS and $\sigma = 80\text{ MeV fm}^{-2}$.
Thus the error due to our lack of knowledge of this function is at most about
3\% in small regions of the star.
}
\end{figure}

Even if the effective shear modulus is positive, the lattice can still be
unstable to shear strains if $A_\lat$ is negative.
This is the case for small regions of the lattice
whose effective shear modulus is shown in
Fig.~\ref{mu_eff_d_vs_p_Hy1_sigma80}, specifically, where $d$ is slightly less than 3.
However, one only obtains such an instability when one uses $f^\lat_\mathrm{cos}$,
$f^\lat_\mathrm{inf}$, or something similar for the lattice interpolation
function.
If one uses $f^\lat_\mathrm{sup}$, then $A_\lat$ remains positive.
This is
illustrated in
Fig.~\ref{Alat_vs_p_Hy1_f_lats} for the same Hy1 $\sigma = 80\text{ MeV fm}^{-2}$ case, using the
lattice interpolation function $f^\lat_\mathrm{cos}$ as well as the sup and inf over all
interpolation functions [given in Eqs.~\eqref{f_bounds}].
We do not use $f^\lat_\mathrm{sup}$ itself in our shear modulus calculations because it does not satisfy the
smoothness and convexity criteria, though Fig.~\ref{Alat_vs_p_Hy1_f_lats} indicates that an
$f^\lat$ that was somewhat greater than $f^\lat_\mathrm{cos}$ for $d$ near $3$ would
not lead to the instabilities.
Regardless, the variations in shear modulus are at most 3\% between sup and inf
in small regions, as illustrated in Fig.~\ref{mu_eff_f_lat_ratios}, so we
tolerate this small inconsistency.

The sign issue does not arise with $A_\perp$, which is made positive by the
contributions from changing the cell size and the stressed background [cf.\ Eq.~\eqref{Aperp}].
While $c_{11}$ is always negative, $A_{\perp,\cell}$ is enough to make
$A_\perp$ positive except for $d \simeq 2.99$, where $-P_\es$ or $A_{\perp,Q}$ are also
needed in certain cases. There $A_{\perp,Q}$ is positive, although it is negative for $d \lesssim 2$.
(However, $-P_\es$ is always positive.)

\begin{figure}[htb]
\begin{center}
\epsfig{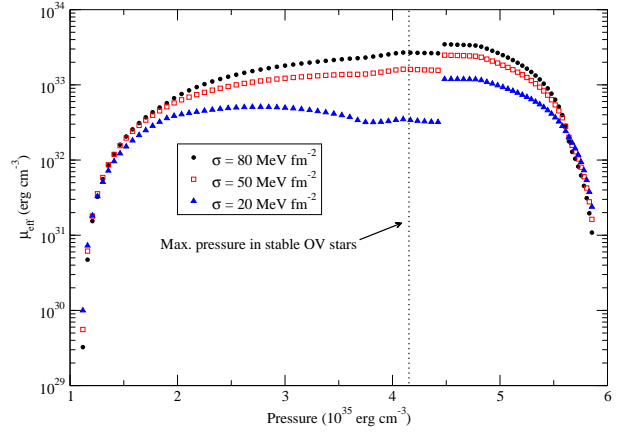}
\end{center}
\caption[$\mu_\eff$ vs.\ pressure for the Hy1 EOS and a variety of surface
tensions]{\label{mu_eff_vs_p_Hy1_sigmas} The (dimensionally continued) 
effective shear modulus versus pressure for the Hy1 EOS and a variety of surface
tensions, $\sigma$.
The discontinuity, which does not occur in a real star, is largely caused by
our simplified treatment of charge screening and serves as an estimate of the
error in that treatment.
For high or moderate surface tensions, that error is less than a factor of 2.
For low surface tension the error can be somewhat larger, but these
lattices are nearly unstable in any case, as shown below.
}
\end{figure}

\begin{figure}[htb]
\begin{center}
\epsfig{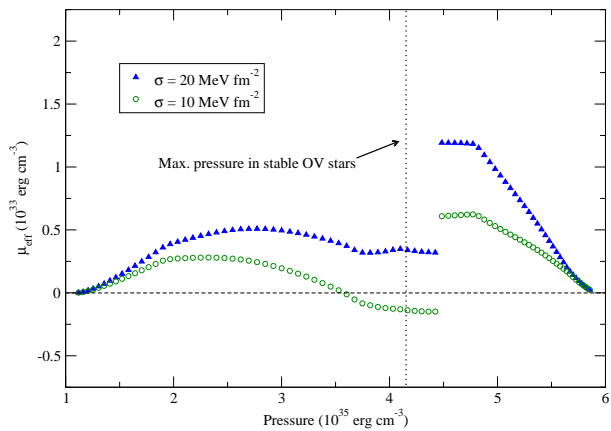}
\end{center}
\caption[$\mu_\eff$ vs.\ pressure for the Hy1 EOS and two low surface
tensions]{\label{mu_eff_vs_p_Hy1_sigma10_20} The (dimensionally continued) 
effective shear modulus versus pressure for the Hy1 EOS and two choices of
$\sigma$, showing the negative values for $\sigma = 10 \text{ MeV fm}^{-2}$.
}
\end{figure}

The largest effect of the equation of state parameters on observable quantities
involving the shear modulus is the extent of the mixed phase in
stable stars. See Weissenborn \emph{et al.}~\cite{Weissenbornetal} for a
survey of the dependence of the mixed phase's extent on the bag constant and
QCD coupling constant for two hadronic EOS parameter sets.

The largest effect
on the shear modulus itself, however, comes from the surface tension, $\sigma$.
This is
illustrated for the Hy1 EOS in Fig.~\ref{mu_eff_vs_p_Hy1_sigmas}. (The magnitude of the reduction of the shear modulus with decreasing surface tension also holds for the
other EOSs we consider.) Note that
the shear modulus is greater for smaller surface tensions for $d$ close to
$3$. Additionally,
 $\sigma$ has a direct, and quite substantial, effect on the lattice's
stability: 
If $\sigma$ is too small, then the lattice will become unstable 
to shears (indicated by $A_\perp$ becoming negative) as the dimensionality
decreases, as illustrated for the Hy1 EOS and two small values of $\sigma$ in
Fig.~\ref{mu_eff_vs_p_Hy1_sigma10_20}.
A surface tension that is much higher than the range we consider makes any
lattice too energetically expensive to form.

\begin{figure}[htb]
\begin{center}
\epsfig{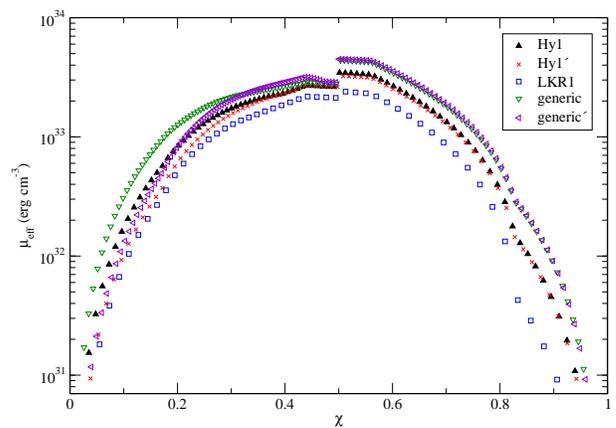}
\end{center}
\caption[$\mu_\eff$ vs.\ $\chi$ for various EOSs]{\label{mu_eff_vs_chi_EOSs} The (dimensionally continued) 
effective shear modulus versus $\chi$ for a representative selection of the
EOSs from
Table~\ref{EOS_table} and $\sigma = 80 \text{ MeV fm}^{-2}$. We have left off
a few points with low shear moduli at either extreme of $\chi$ to better show
the differences between the different predictions where the shear moduli are
the highest and most astrophysically relevant.
There the uncertainty in shear modulus due to factors other than surface
tension is at most about a factor of $2$.
}
\end{figure}

To illustrate the relatively small effect of the other EOS parameters on the effective
shear modulus for a fixed surface tension, we plot $\mu_\eff$ versus
the quark volume fraction $\chi$ for a representative sample of the EOSs from
Table~\ref{EOS_table} with $\sigma = 80\text{ MeV fm}^{-2}$ in
Fig.~\ref{mu_eff_vs_chi_EOSs}. The different flavors of Hy$1$ EOS---i.e.,
Hy$1$, Hy$1\mu$, Hy$1\sigma$, Hy$1\mu\sigma$---all have quite similar shear
moduli for a fixed surface tension. The inclusion of the surface tension
contribution to the pressure balance causes the largest difference of any of these other EOS
parameters, but even this difference is
considerably smaller than the differences between the EOSs shown in the figure,
and is negligible where the shear modulus is largest. We have thus not shown
the traces for those EOSs, to avoid a cluttered figure.

It is interesting to compare these shear modulus values to those for the lattice of nuclei in the neutron star
crust as well as those for crystalline color superconducting (CSC) quark matter:
The crustal shear modulus ranges from $\sim 6\times 10^{27}$ to $\sim2\times 10^{30} \text{ erg cm}^{-3}$ (see, e.g., Fig.~2 in~\cite{HJA}), while the shear modulus of CSC quark matter computed by Mannarelli, Rajagopal, and Sharma (MRS)~\cite{MRS} could be as large
as $\sim4\times10^{34}\text{ erg cm}^{-3}$; the lower bound they give is $\sim 8\times10^{32}\text{ erg cm}^{-3}$.
The hadron--quark mixed phase shear moduli we have computed range from $\sim 10^{30}\text{ erg cm}^{-3}$ for small blobs to $\sim 4\times10^{33}\text{ erg cm}^{-3}$ for hadronic slabs (with a surface tension of $\sigma = 80 \text{ MeV fm}^{-2}$).
Thus the mixed-phase shear modulus is at least comparable to the largest shear
modulus in the crust and can be three orders of magnitude higher than it, and
the largest mixed-phase shear modulus is an order of magnitude less than the
largest CSC quark matter shear modulus, roughly the geometric mean of the
range given by MRS.
(Note, however, that the maximum mixed-phase shear modulus for a given set of
EOS parameters is not necessarily realized in a stable star.)

\section{Conclusions}
\label{concl1}

We have made a more careful calculation of the shear modulus of the
hadron--quark mixed phase than has previously been attempted~\cite{OwenPRL,
Nayyar, HPY}. In particular,
we have computed all the lattice's (anisotropic) shear elastic constants, before
averaging to
obtain an isotropic effective shear modulus for a polycrystal.
We have also dealt with the lattice's changing dimension in both the electrostatic
potential and the geometrical effects on the elastic constants. Perhaps most
importantly, we have included the contributions to the elastic constants from
changing the size of the blobs for $d < 3$. These act to stabilize the lattice
for lower dimensions (leading to significant contributions to the shear modulus
from these portions of the lattice), though only for sufficiently large surface
tensions.
We have found that for our choices of parameters, most of the lower-dimensional portions of the
lattice are unstable to shear perturbations if the surface tension is less
than 10--20~MeV\,fm$^{-2}$. (As discussed in~\cite{HPS, Alfordetal_PRD}, the mixed phase is
not favored if the surface tension is too large. However, as we mention in Sec.~\ref{lat_struc}, and will
explore in depth in~\cite{paper2}, these are local energy considerations, while the mixed phase
may still be favored by global energy considerations.)

These calculations depend upon a wide variety of poorly constrained parameters.
While we found that the shear modulus of the mixed phase itself depends most
sensitively on the surface tension, astrophysical effects depend on
the amount of mixed phase present in a given star, which is primarily
determined by
the standard hybrid EOS parameters. We will see this
in more detail when we
use the shear moduli calculated here to compute the maximum elastic quadrupolar
deformations of these hybrid star models in a companion paper~\cite{paper2}.

The obvious place where the shear modulus calculation could be improved significantly is the
treatment of charge screening: We know from the calculations of
Endo~\emph{et al.}~\cite{Endoetal, Endo} that including (nonlinear) charge screening in the
computation of the cell energy leads to
significant differences in the lattice properties at a given density.
Here one would like to at least use the Thomas-Fermi
approximation, if one did not perform a nonlinear
calculation as in~\cite{Endoetal, Endo}. The shear modulus will be affected both by the change
in cell size and spacing, as
well as through the cell energy's direct contribution to one of the elastic
constants for lower dimensions.
Since this contribution is necessary to stabilize the lattice for those dimensions, it would
be interesting to see whether our discovery that the lattice is only
stable for sufficiently large surface tensions still holds with
charge screening.
For instance, there will be, in effect, Fermi contributions to the cell energy
in a proper treatment.

One might also want to investigate the effects of using different descriptions
of the hadronic and quark
matter.
For instance, for the quark matter, one could use the higher-order perturbative calculations of Kurkela,
Romatschke, and Vuorinen~\cite{KRV} or the Nambu--Jona-Lasinio treatment
(e.g.,~\cite{Blaschkeetal, BoSe}), while Dirac-Brueckner-Hartree-Fock
(e.g.,~\cite{Blaschkeetal}) would be a possibility for the hadronic matter. One might also investigate the potential effects of color superconductivity (see~\cite{Alford2008} for a review): The inclusion of
CSC quark matter would, of course, increase the shear modulus (see Mannarelli, Rajagopal, and Sharma~\cite{MRS} for a calculation of the shear modulus of bulk CSC quark matter). However, if one does not have a crystalline phase, color superconductivity would primarily affect these shear modulus calculations through the EOS, where Alford~\emph{et al.}~\cite{Alfordetal} find that including color superconductivity reduces the transition density to quark matter, but does
not appreciably change the maximum mass. While one would also expect color superconductivity to affect the
energy of the quark blobs, and thus the contributions to the shear modulus from changing the blob size, we have not even included quark interaction contributions to the blob energy in the present calculation, only considering the electrostatic contributions. Other possibilities for extending the calculation overall would be
including more exotica (hyperons, for instance, as in, e.g.,~\cite{CBBS, BoSe, MCST}), and, in particular, magnetic fields [cf.\ the discussion
in Baiko~\cite{Baiko} and below our Eq.~\eqref{mu_eff1}].

One could also apply our methods of calculating the shear modulus of the pasta phases
to the nuclear pasta appearing in the crust. Such a calculation would be particularly interesting
given recent studies of the observational effects of the crustal pasta which used very rough
models for its shear modulus~\cite{Sotani, GNHL}.
Our methods could also be adapted to calculate the shear moduli of meson
condensates more carefully than the existing order of magnitude estimates
(Ref.~\cite{HPY} and references therein).

The shear moduli here can be several hundred times larger than those first
estimated~\cite{OwenPRL}.
Na{\"\i}vely, one might expect the maximum quadrupole to go up by a similar factor,
but this is complicated by issues of where the various lattices occur in the
star.
We will present an exploration of that and other issues with the maximum quadrupole
elsewhere~\cite{paper2}.

\acknowledgments

We are grateful to M.\ Alford and C.\ Horowitz for helpful discussions.
This work was supported by NSF grants PHY-0555628 and PHY-0855589, the
Eberly research funds of Penn State, and the DFG SFB/Transregio 7.

\appendix*

\section{Checks of lattice sums}
\label{checks}

We have checked that our code can reproduce all the relevant results for
elastic constants presented in the literature, and detail these checks here:
For a three-dimensional unscreened bcc lattice, we have checked that
we can reproduce the results of Fuchs~\cite{Fuchs} (for $c_{44}$ and $A_\lat$)
and Ho~\cite{Ho} (for $c_{11}$), and also that we are in agreement with the very recent
and much more precise calculation of all three elastic constants (to $8$
significant digits) by Baiko~\cite{Baiko}. For a three-dimensional bcc lattice
with screening,
we have checked against the results from Horowitz and Hughto~\cite{HH}. We have
also checked that the surprisingly simple results we obtain in the unscreened
two-dimensional hexagonal case agree with those obtained analytically using zeta function
regularization.

Numerical agreement is good.
We reproduce the Fuchs results to better than $0.25\%$ for $A_\lat$ and
$0.015\%$ for $c_{44}$.
Moreover, we agree with Baiko's results to all five significant figures to
which we have calculated them (using a screening length of $10^{10}a$).
Baiko claims agreement with Fuchs and does not comment on the discrepancy,
which we conjecture is due to modern computational technology allowing for the
summation of more terms.
We agree with Ho's result to the four significant figures to which he gives it.
Horowitz and Hughto computed $A_\lat $ (twice their $b_{11}$) for a screening length of $\lambda = 0.863a$.
We agree to better than $1\%$ for $A_\lat$ and $0.011\%$ for $c_{44}$.
The discrepancy for $c_{44}$ could be due to rounding.
The discrepancy for $A_\lat$ is compatible with Horowitz and Hughto's caveats
about a 1\% error in that part of their calculation due to using a brute force
summation rather than the Ewald method.\footnote{N.B.: Horowitz and Hughto's
expression for $b_{11}$ in terms of $c_{11}$ and $c_{12}$ in their Eq.~(13) is missing a factor of $1/2$, which can be seen to be present
by starting from Eq.~(7) in~\cite{OI} and noting that $c_{31} = c_{12}$ for a cubic lattice.}

For the two-dimensional unscreened hexagonal lattice, we obtain what appear
numerically to be simple fractions for the elastic constants (when expressed
in Fuchs-type units of $Q^2$).
Explicitly, we have $A_\lat = 1/2$, $c_{11} = -1/4$, and $c_{44} = 1/4$. 

We can show that these elastic constants indeed take the expected, extremely
simple forms by calculating them analytically using analytic continuation of
an Epstein zeta function, as discussed in, e.g.,~\cite{CG, GZ}. (A calculation of these elastic
constants for a $2$-dimensional Coulomb lattice does not
appear to exist in the literature.)
We first note that the $2$-dimensional Coulomb potential is $-2Q\log
r$.
Thus the sums that give the elastic constants will all have the formal (and
divergent) form
\<
\sideset{}{'}\sum_{\vec{x}\in\Lambda_\hex}\alpha,
\?
where $\alpha$ is some constant and $\Lambda_\hex$ is the hexagonal lattice.
To see this explicitly for $c_{11}$, we have (writing it in Fuchs-type units)
\<
c_{11}^\hex = -\sideset{}{'}\sum_{\vec{x}\in\Lambda_\mathrm{hex}} \left(\frac{x_1^2}{\|\vec{x}\|^2} - 2\frac{x_1^4}{\|\vec{x}\|^4}\right) = \sideset{}{'}\sum_{\vec{x}\in\Lambda_\mathrm{hex}}\frac{1}{4},
\?
where we have used the substitutions $x_1^2 \to \|\vec{x}\|^2/2$ and $x_1^4 \to (3/8)\|\vec{x}\|^4$ (valid when
summing over a hexagonal lattice). (One can
obtain these substitutions by direct calculation, for which it is convenient to
write the lattice points as elements of $\C$, so one can multiply a given point
by sixth roots of unity to obtain all the other lattice points the same
distance from the origin.)

Of course, this sum diverges, but, as indicated in
Sec~\ref{elastic_calc},
we expect to obtain a regularized version from
our Ewald sum procedure. Indeed, it seems reasonable (possibly even likely)
that we should obtain the
same result using Ewald sums as one obtains using zeta function regularization, since both methods rely at their
root on the Jacobi imaginary transformation of theta functions: See, for instance, the discussion in~\cite{NKJ-M},
which obtains the Poisson summation formula used in the Ewald method from the Jacobi transformation, and the discussion
in Appendix~B of~\cite{CG}, which analytically continues the Epstein zeta function using the Jacobi imaginary transformation of theta functions
(referred to by its alternate name of the modular property of the theta function).

We thus begin by defining the Epstein zeta function $\zeta_\Lambda$ associated
with an arbitrary lattice $\Lambda$. This is given [cf.\ the version for $\Z^d$ in
Eq.~(B1) of~\cite{CG} and the more general version in Eq.~(2.5) of~\cite{GZ}~]
by
\<
\zeta_\Lambda(s) := \sideset{}{'}\sum_{\vec{x}\in\Lambda}\frac{1}{\|\vec{x}\|^{2s}}
\?
for $\Real s > d/2$ (where $d$ is the dimension of $\Lambda$), and analytically continued to $s\in\C$ (except for
isolated poles) by Riemann's integral-splitting procedure and the Jacobi imaginary transformation, as
in~\cite{CG,GZ}. We have $\zeta_\Lambda(0) = -1$ for any lattice [of positive dimension---cf.\ the third
property given in Appendix~B of~\cite{CG} and Eq.~(2.13) in~\cite{GZ}~], so we obtain $c_{11} = -1/4$, in Fuchs-type
units, as advertised above. We obtain the values for $c_{44}$ and $A_\lat$ given above by the same procedure, except that
we also need to use the additional hexagonal lattice sum substitution $x_1^2x_2^2 \to \|\vec{x}\|^4/8$.

\bibliography{paper1}

\end{document}